\documentclass[aps,prb,twocolumn,notitlepage]{revtex4-2}
\usepackage{lipsum}
\usepackage{multirow}
\usepackage{amsmath,graphicx,color}
\usepackage{bm}
\usepackage{natbib}
\usepackage{placeins}
\setcitestyle{square}
\usepackage{braket}
\usepackage{makerobust}
\usepackage{xcolor}
\usepackage{hyperref}
\hypersetup{%
 colorlinks,
 breaklinks=true,
 plainpages=false,%
 citecolor=blue,
 linkcolor=blue,
 urlcolor=blue,
 bookmarksopen=true,%
 bookmarksnumbered=false,%
 bookmarksdepth=5%
}

\newcommand{\makeauthor}[2]{\newcommand{#1}[1]{{%
 \sffamily\color{#2}{%
 \bfseries\begingroup\escapechar=-1\edef\x{\endgroup\string#1}\x:%
 } ##1}}%
 \MakeRobustCommand#1}
\makeauthor{\dkm}{red}
\makeauthor{\rew}{blue}
\makeauthor{\jb}{purple}
\begin{document}

\title[Article Title]{Finding Hidden Numbers with Majorana-based Topological Quantum Algorithms:\\
Simulation of the Bernstein-Vazirani Algorithm}

\author{Jasmin Bedow and Dirk K. Morr}
\affiliation{Department of Physics, University of Illinois Chicago, Chicago, IL 60607, USA}

%\date{\today}

\begin{abstract}
 \bf
 Executing quantum algorithms using Majorana zero modes - a major milestone for the field of topological quantum computing - requires a platform that can be scaled to large quantum registers, can be controlled in real time and space, and a braiding protocol that uses the unique properties of these exotic particles. Here, we demonstrate the first successful simulation of the Bernstein-Vazirani algorithm in two-dimensional magnet-superconductor hybrid structures from initialization to read-out of the final many-body state.  Utilizing the Majorana zero modes' topological properties, we introduce an optimized braiding protocol for the algorithm and a scalable architecture for its implementation with an arbitrary number of qubits. We visualize the algorithm protocol in real time and space by computing the non-equilibrium density of states, which is proportional to the time-dependent differential conductance, and the non-equilibrium charge density, which assigns a unique signature to each final state of the algorithm.
\end{abstract}

\maketitle

%%%%%%%%%%%%%%%%%%%%%%%%%%%%%%%%%%%%%%%%%%%%%%%%%%%%%%%%%%
%%%%%%%%%%%%%%%%%%%%%%%%%%%%%%%%%%%%%%%%%%%%%%%%%%%%%%%%%% }

{\it Introduction.~}
Topological superconductors harbor Majorana zero modes (MZMs), which have been proposed as a promising platform for the realization of a fault-tolerant quantum computing due to their non-Abelian braiding statistics and robustness against disorder and decoherence effects \cite{Nayak_2008, Sarma_2015, Beenakker_2019}. Magnet-superconductor hybrid (MSH) systems, consisting of magnetic adatoms placed on the surface of $s$-wave superconductors, have been shown to be a suitable material platforms for the design of topological superconducting phases, providing strong evidence for the existence of MZMs in one- \cite{Nadj-Perge_2014, Ruby_2015, Pawlak_2016,Kim_2018} and two-dimensional \cite{Palacio-Morales2019,Menard2017,Kezilebieke2020,Bazarnik2023} systems. To theoretically test the feasibility of MZMs for the implementation of topologically protected quantum algorithms in MSH systems, it is necessary to translate established quantum algorithms \cite{Deutsch_1992, Berthiaume_1992, Bernstein_1997, Grover_1996, Shor_1999} into braiding protocols for Majorana zero modes \cite{Kraus_2013}, and successfully simulate these in real time and space. While this has been achieved for the Clifford gates -- the fundamental building blocks of topologically protected quantum algorithms --
in $N=2$ qubit systems on various material platforms \cite{Alicea_2011,Halperin_2012,Sekania_2017,Harper_2019,Tutschku_2020,Mascot_2023,Hodge_2025,Kraus_2013,Amorim_2015,Aasen_2016,Karzig_2017,Zhou_2022,Sanno_2021,Hyart_2013,Li_2016,Bedow2024}, the crucial extension to more complex quantum algorithms and larger ($N>2$) qubit systems has not yet been considered. One of the most important quantum algorithms, demonstrating a polynomial speed-up over classical algorithms, is the Bernstein-Vazirani (BV) algorithm \cite{Berthiaume_1992, Bernstein_1997}, in which a hidden number $s$, encoded into an oracle function $U_s$, can be read-out from the final state of an $N$ qubit system after a single query of the oracle. While experimental attempts to implement the BV algorithm in non-topological systems \cite{Du_2001,Brainis_2003,Londero_2004,Peng_2004,Debnath_2016,Wright_2019} have been reported, no demonstration of its feasibility and scalability in topologically protected, MZM-based systems, has been provided to-date.

In this article, we demonstrate the successful simulation of the BV algorithm from initialization to read-out using MZMs in a $N=3$ qubit MSH system. To this end, we combine the recent proposal for MZM braiding via the manipulation of the magnetic structure in MSH systems -- motivated by electron-spin resonance - scanning tunneling microscopy (ESR-STM) experiments, demonstrating that individual spins in small magnetic clusters can be flipped \cite{Yang_2018, Yang_2019, Phark_2023, Wang_2023} -- with recent theoretical advances in simulating time-dependent phenomena in superconductors
%without an exponentially scaling Hilbert space
\cite{Bedow2024, Mascot_2023, Hodge_2025}. This combination enables us to simulate the BV algorithm and compute its success probabilities on experimentally relevant time and length scales. We show how the choice of a network architecture allows us to initialize the system in various pure $N$-qubit many-body states both in the even- and odd-parity sector. Moreover, we demonstrate that the hidden number $s$ can be experimentally read out from the spatial form of the charge density after fusing MZM pairs at the conclusion of the BV algorithm. Finally, we visualize the entire BV algorithm in time and space by using the time-, energy- and spatially-resolved non-equilibrium density of states, $N_{neq}$ \cite{Bedow_2022} to image the MZM world lines. Our results establish for the first time the feasibility to implement topologically protected quantum algorithms in MSH systems, providing an important step towards the
realization of topological quantum computing.

{\it Theoretical Methods.~} As a platform for the implementation of the BV algorithm, we use MSH structures, consisting of a network of magnetic adatoms placed on the surface of a two-dimensional (2D) superconductor, which are described by the Hamiltonian
\begin{align}
\mathcal{H} =& \; -t_e \sum_{{\bf r}, {\bf r}', \beta} c^\dagger_{{\bf r}, \beta} c_{{\bf r}', \beta} - \mu \sum_{{\bf r}, \beta} c^\dagger_{{\bf r}, \beta} c_{{\bf r}, \beta} \nonumber \\
 &+ \mathrm{i} \alpha \sum_{{\bf r}, {\bm \delta }, \beta, \gamma} c^\dagger_{{\bf r}, \beta} \left({\bm \delta} \times \boldsymbol{\sigma} \right)^z_{\beta, \gamma} c_{{\bf r} + {\bm \delta}, \gamma} \nonumber \\
 & + \Delta \sum_{{\bf r}} \left( c^\dagger_{{\bf r}, \uparrow} c^\dagger_{{\bf r}, \downarrow} + c_{{\bf r}, \downarrow} c_{{\bf r}, \uparrow} \right) \nonumber \\
 &+ J {\sum_{{\bf R} , \beta, \gamma}} c^\dagger_{{\bf R}, \beta} \left[ {\bf S}_{\bf R}(t) \cdot \boldsymbol{\sigma} \right]_{\beta,\gamma} c_{{\bf R}, \gamma} \; .
 \label{eq:H}
\end{align}
Here, $c^\dagger_{{\bf r}, \beta}$ creates an electron with spin $\beta$ at site ${\bf r}$, $t_e$ denotes the nearest-neighbor hopping parameter on a 2D square lattice, $\mu$ is the chemical potential, $\alpha$ is the Rashba spin-orbit coupling between nearest-neighbor sites ${\bf r}$ and ${\bf r} +{\bm \delta}$, $\Delta$ is the $s$-wave superconducting order parameter, $J$ is the magnetic exchange coupling between the magnetic adatom with spin ${\bf S}_{\bf R}$ at site ${\bf R}$ and the conduction electrons, and $\boldsymbol{\sigma}=(\sigma_x,\sigma_y,\sigma_z)^T$ is the vector of the Pauli matrices. As the hard superconducting gap suppresses Kondo screening \cite{Balatsky_2006,Heinrich_2018}, we consider the magnetic adatoms as classical spins, whose orientation at a time $t$ is given by $S_{\bf R}(t)$.

Below we choose parameters such that (i)
the MSH network is topological when the magnetic adatom moments possess a ferromagnetic (FM)
out-of plane alignment, but is trivial when the moments exhibit an antiferromagnetic (AFM)
in-plane order, and (ii) the MZM localization length is significantly smaller than the network size we can consider computationally. To move the MZMs in space, we rotate the magnetic moments between the FM and AFM alignments, with the MZMs being always located at the end of the topological (FM) regions. This rotation of magnetic moments, which in principle can be achieved using ESR-STM techniques \cite{Yang_2018,Yang_2019,Wang_2023,Phark_2023} (for a detailed discussion, see Ref.\cite{Bedow2024}), is characterized by a rotation time $T_R$ to rotate a single moment by $\pi/2$ between in- and out-of-plane alignment. To ensure the adiabaticity of the entire BV algorithm, we choose a rotation time $T_R \gg \hbar/\Delta_t$ where $\Delta_t$ is the topological gap in the system (for details, see Methods Section). Below all times are given in units of $\tau_e = \hbar/t_e$ such that for typical values of $t_e$ of a few hundred meV, $\tau_e$ is of the order of a few femtoseconds.

\begin{figure*}[ht]
 \centering
 \includegraphics[width=17cm]{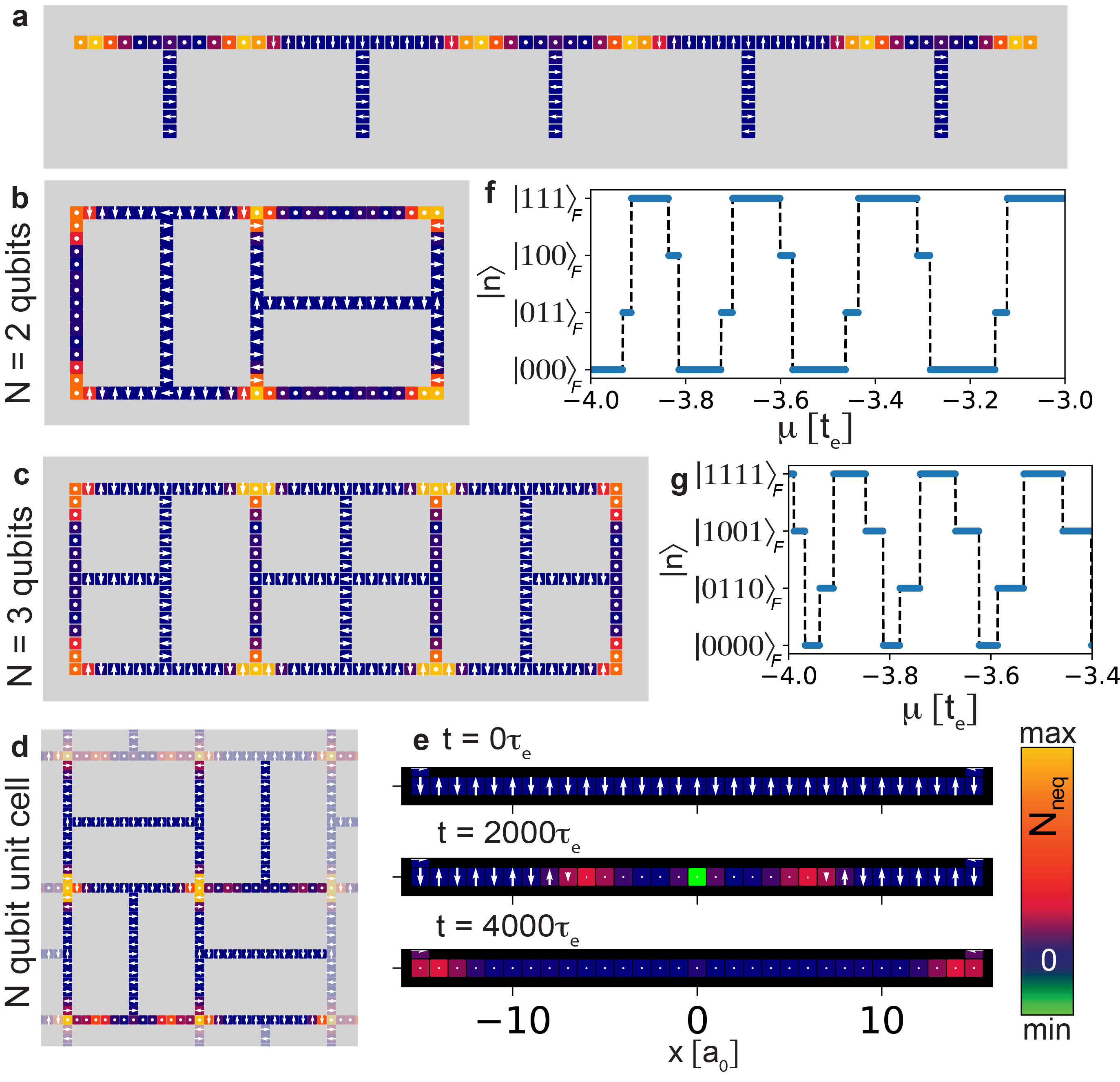}
 \caption{\textbf{Gate Architecture and Initialization.}
 {\bf a}-{\bf c} MSH architectures for the simulation of the Bernstein-Vazirani algorithm. Spin orientations are represented by white arrows (white dots representing an out-of-plane orientation), and the zero-energy $N_{neq}$ (color scale) reveals the existence of MZMs at the end of the topological (FM) rungs.
 {\bf d} Unit cell (shown in full color, adjacent unit cells are shown in opaque) to generalize the MSH architecture to a (large) $N$ qubit system. {\bf e} $N_\mathrm{neq}$ for several times during the initialization process along a single rung in the MSH structure.
 Initialized state $\ket{n}$ as a function of the chemical potential for the {\bf f} $N=2$ and {\bf g} $N=3$ qubit systems.
 Parameters are $(\mu, \alpha, \Delta, JS) = (-3.93, 0.45, 1.2, 2.6) t_e$, $\Gamma = 0.03 t_e$ with an execution time of $4000 \tau_e$ for the initialization process.}
 \label{fig:Fig1}
\end{figure*}

For an $N$ qubit system, the BV algorithm \cite{Berthiaume_1992, Bernstein_1997} encodes a hidden number $s=0,...,2^N-1$ in an oracle function $U_s$, which is translated into a specific braiding protocol of the MZMs. The time evolution of the entire system's many-body wave-function $\ket{\Psi(t)}$ under the braiding protocol is computed using a recently developed non-equilibrium formalism \cite{Bedow2024, Mascot_2023} (for details, see Methods Section). Starting point for all our simulations is the topologically trivial, even-parity ground state of the system, $\ket{GS}$, in which all spins of the MSH network are aligned AFM in-plane; thus $\ket{\Psi(t=0)}=\ket{GS}$. By changing the spin orientation in certain segments of the network from in-plane AFM to out-of-plane FM, the system becomes topological and is initialized into one of the $2^N$ pure (many-body) qubit states $\ket{n}$ at time $t=t_i$. The application of the BV algorithm as reflected in the braiding protocol then yields $\mathrm{BV}_{s} \ket{n} = \ket{n \oplus s}$. To evaluate the success of the BV algorithm, we define the time-dependent transition probability $p_{n}(t)=|\langle \Psi(t) | n \rangle|^2$. Thus, $p_{n}(t_i)=1$ describes the successful initialization of the system into a desired initial state $\ket{n}$, and $p_{n \oplus s}(t_c) = |\langle \Psi(t_c) | n \oplus s \rangle|^2 =|\langle \Psi(t_c) | BV_s | n \rangle|^2=1$ reflects a perfect implementation of the BV algorithm which concludes at time $t=t_c$. While a classical computer requires $N$ queries of the oracle to find the hidden number $s$ with absolute certainty, a quantum computer requires a single query, being able to obtain $s$ from a comparison of the initial $\ket{n}$ and final $\ket{n \oplus s}$ states. As we show below, the read-out of these states can be experimentally achieved by measuring the local charge density (for details, see Methods Section)
\begin{equation}\label{eq:charge_dense}
 \rho({\bf r},\sigma,t) = -e \braket{\Psi(t)|c^\dagger_{{\bf r}, \sigma} c_{{\bf r}, \sigma}|\Psi(t)} \;
\end{equation}
after fusion of MZM pairs, as it directly reflects a pair's occupation, and thus allows to uniquely identify all $2^N$ many-body states of an $N$ qubit system.

Finally, we visualize the entire BV algorithm from initialization to fusion using the energy-, time- and spatially-resolved non-equilibrium density of states $N_{neq}({\bf r}, \sigma, t, \omega)$, which is proportional \cite{Bedow_2022} to the time-dependent differential conductance $dI(V,{\bf r},t)/dV$ measured in STM experiments (for details, see Methods Section). \\

\begin{table*}
 \centering
 \begin{tabular}{|c|c|c|c|c|c|c|c|c|c|c|c|c|c|c|}
 \hline
 Fock State & \multirow{2}{.7cm}{N=2}& {$\ket{001}_F$} & {$|010\rangle_F$}& {$|100\rangle_F$} & {$|111\rangle_F$} &\multirow{2}{0.7cm}{N=3} &{$|0110\rangle_F$} & {$|1010\rangle_F$} & {$|1100\rangle_F$} & {$|0000\rangle_F$} & {$|0011\rangle_F$} & {$|1001\rangle_F$} & {$|1111\rangle_F$} & {$|0101\rangle_F$} \\
 \cline{1-1} \cline{3-6} \cline{8-15}
 Qubit State & & {$|00\rangle_q$} & {$|01\rangle_q$} & {$|10\rangle_q$} & {$|11\rangle_q$}& & {$|011\rangle_q$} & {$|010\rangle_q$} & {$|001\rangle_q$} & {$|000\rangle_q$} & {$|111\rangle_q$} & {$|101\rangle_q$} & {$|110\rangle_q$}
 & {$|100\rangle_q$} \\
 \hline
 \end{tabular}
 \caption{Correspondence between states in the Fock occupation notation $\ket{...}_F$ and the logical qubit notation $\ket{...}_q$.}
 \label{tab:fock_qubit_conversion}
\end{table*}

\begin{figure*}
 \centering
 \includegraphics[width=17cm]{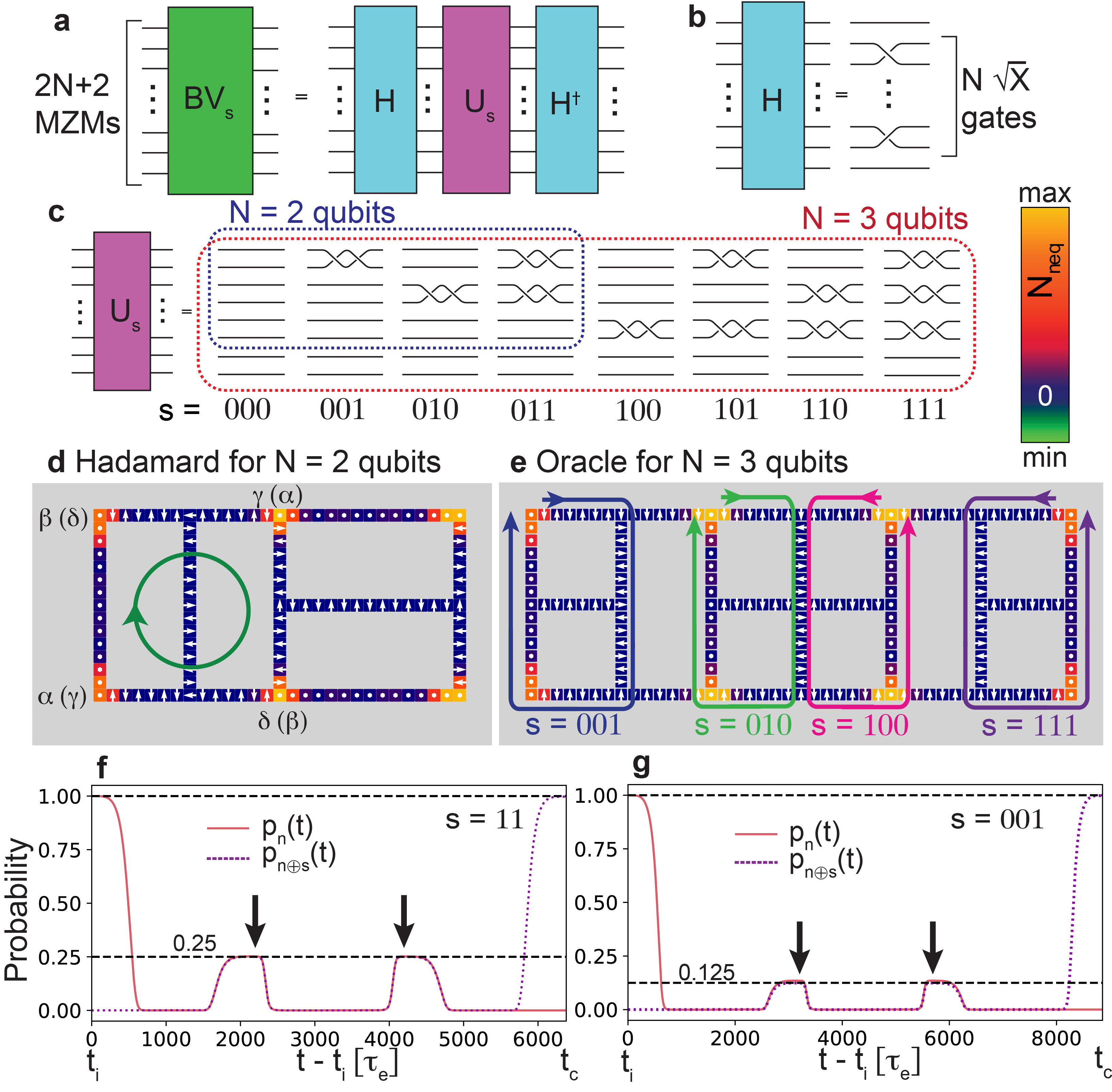}
 \caption{\textbf{Bernstein-Vazirani Algorithm.}
 {\bf a} Schematic of the Bernstein-Vazirani algorithm for $N$ qubits (with $2N+2$ MZMs). Braiding protocol for {\bf b} the $N$ qubit Hadamard gate using $N$ $\sqrt{X}$-gates, and {\bf c} the oracle function $U_s$ for $N=2,3$ qubits using $Z$-gates.
 Real space braiding processes for {\bf d} the Hadamard gate for a $N=2$ qubit system, and {\bf e} the oracle function for $N=3$ and $s \in \{001,010,100,111\}$. Time-dependent transition probabilities for the {\bf f} $N=2$ qubit system with $s=11$, and the {\bf g} $N=3$ qubit system with $s=001$, starting from the initialization time $t_i$.
 Parameters are $\mu = -3.913 t_e$ ($-3.93 t_e$), $(\alpha, \Delta, JS) = ( 0.45, 1.2, 2.6) t_e$ with execution times of $2200 \tau_e$ ($3200 \tau_e$) for the Hadamard gate and $1980 \tau_e$ ($2460 \tau_e$) for the oracle for the $N=2$ ($N=3$) qubit system.}
 \label{fig:Fig2}
\end{figure*}

{\it Results.~} To identify a hidden number $s$ using the Bernstein-Vazirani algorithm, one proceeds in three steps: (i) initialization of the system in a many-body state $\ket{n}$, (ii) execution of the BV algorithm as represented by a MZM braiding protocol,
(iii) read-out of the final state $\ket{n \oplus s}$. Below, we will discuss each step separately to elucidate how the system's architecture affects its initialization and the implementation of the BV algorithm.\\

{\it Gate architecture.~}
While T-structures (see Fig.~\ref{fig:Fig1}\textbf{a}) have previously been discussed as the basic computing architecture \cite{Alicea_2011}, it was recently proposed that more compact, folded MSH networks (see Figs.~\ref{fig:Fig1}\textbf{b}, \textbf{c}) provide several advantages as they enable additional braiding operations and the use of spatial symmetries to efficiently implement quantum gates \cite{Bedow2024}. Below, we therefore consider the network architectures shown in Figs.~\ref{fig:Fig1}\textbf{b}, \textbf{c} for the simulation of the BV algorithm in $N=2$ and $3$ qubit systems, respectively. These structures can be generalized to large $N$ qubit systems by using the unit cell structure shown in Fig.~\ref{fig:Fig1}\textbf{d} as the building block.\\

{\it Initialization.~}
The initialization of the system into a many-body qubit eigenstate $\ket{n}$
starts from the even-parity ground state of the trivial phase, denoted by $\ket{GS}$, in which the spins possess an AFM in-plane alignment. To create three (four) topological rungs in the MSH network of Fig.~\ref{fig:Fig1}\textbf{b} (Fig.~\ref{fig:Fig1}\textbf{c}), each of which exhibits two MZMs at its ends, we rotate the spins consecutively to an out-of-plane ferromagnetic (FM) alignment, starting in the center of the rung. The non-equilibrium LDOS for one of the rungs for several times during this initialization process is shown in Fig.~\ref{fig:Fig1}\textbf{e}. The resulting many-body eigenstate of the MSH network in the Fock notation is denoted by
$\ket{n}=\ket{l_1l_2 ...}_F$, with $l_i \in \{0,1\}$
describing the occupation of the MZM pair on the $i^{th}$ rung,
and depends not only on the system's chemical potential, $\mu$, but also on the specific MSH architecture as shown in Figs.~\ref{fig:Fig1}\textbf{f} and \textbf{g} for the $N=2$ and $N=3$ qubit architecture, respectively.
While the initialization in all cases starts from $\ket{GS}$, an occupation of an individual Majorana pair different from zero occurs when the ferromagnetic moment of the initialized topological rung is sufficiently strong to break a Cooper pair, resulting in a local phase transition that changes the parity of the superconducting ground state. This is similar to the local phase transition that an s-wave superconductor undergoes when the scattering strength of a single \cite{Sakurai_1970, Salkola_1997, Bazaliy_2000, Balatsky_2006} or multiple magnetic impurities \cite{Morr_2003a,Morr_2006a} exceeds a critical value.
Due to the spatial symmetry of the $N=3$ qubit system (Fig.~\ref{fig:Fig1}\textbf{g}), two of the four rungs simultaneously undergo a phase transition and thus change their occupation in the initialization process, implying that the possible many-body states after initialization are always in the even-parity sector. In contrast, for the $N=2$ qubit system (Fig.~\ref{fig:Fig1}\textbf{f}) only two of the three rungs possess the same spatial symmetry, and hence the many-body states after initialization can either be in the even- or odd-parity sector, depending on whether an even (0 or 2) or odd (1 or 3) number of rungs undergoes the phase transition. Thus, by starting from a single trivial state $\ket{GS}$ and utilizing the spatial symmetry of the MSH architecture, one can initialize the system into specific initial topological many-body states $\ket{n}$, both in the even- and odd-parity sector.\\

{\it BV algorithm.~}
The BV algorithm is decomposed into a Hadamard gate that creates a superposition of all qubit states, an Oracle function that encodes the hidden number $s$, and a second Hadamard gate that reverses the superposition of qubit states (see Fig.~\ref{fig:Fig2}{\bf a}). We generalize the Hadamard gate for an $N$ qubit system with $2N+2$ MZMs by employing $N$ $\sqrt{X}$ gates as sketched in Fig.~\ref{fig:Fig2}{\bf b} (see Supplementary Note 1). Here, the $\sqrt{X}$-gate ($\sqrt{Z}$-gate) represents a braid of two MZMs from adjacent pairs (the same pair). This Hadamard gate results in an equal-weight superposition of all $2^N$ pure $N$-qubit states involving complex phases (see Supplementary Note 1) and is thus non-Hermitian $H \neq H^\dagger$. We therefore need to employ $H^\dagger$ as the second Hadamard gate in the BV algorithm, which is realized by inverting the directions of the MZM braids in $H$. We note that this representation of the Hadamard gate not only significantly reduces the number of required braiding operations and allows for their simultaneous execution, but also realizes a unique generalization to $N \geq 2$ qubit systems, in contrast to previous proposals using combinations of $\sqrt{X}$- and $\sqrt{Z}$-gates \cite{Kraus_2013,Beenakker_2019} (see Supplementary Note 1). Finally, we note that the braiding protocol for the oracle function $U_s$ satisfies the generalized form $U_s \ket{n} = (-1)^{n \cdot \hat{A} \cdot s} \ket{n}$, where the matrix $\hat{A}$ is determined by the specific realization of the Hadamard gate (for details, see Supplementary Note 1).

The folded MSH networks shown in Figs.~\ref{fig:Fig1}{\bf b} and {\bf c} (in contrast to the conventional T-structure) allow for an efficient implementation of the MZM braiding protocols shown in Figs.~\ref{fig:Fig2}{\bf b} and {\bf c}.
In particular, the Hadamard gate for the $N=2$ qubit system can be implemented by rotating 4 MZMs by $180^\circ$
clockwise (see Fig.~\ref{fig:Fig2}{\bf d}) -- thus simultaneously realizing two $\sqrt{X}$ gates -- while the $Z$-gates for the oracle are implemented by moving an MZM pair along a plaquette (see Fig.~\ref{fig:Fig2}{\bf e}). Both of these implementations allow for faster execution times as they preserve the distance between the MZMs.

The mapping $\mathrm{BV}_{s} \ket{n} = \ket{n \oplus s}$ holds both in the Fock (occupation) notation ($\ket{...}_F$), as well as in the logical qubit state notation ($\ket{...}_q$), with the value of $s$ given in the respective notation. However, there exists some freedom in choosing the translation between these two notations. Here, in the even-parity sector for $N=3$, we set $\ket{0000}_F=\ket{000}_q$, and a value of "1" in the $i^{th}$ position of $s$ is assigned when a $Z$-gate is performed on the $i^{th}$ pair of MZMs (unless otherwise stated, $s$ is given in the logical qubit notation below). This completely determines the mapping between these two notations shown in Table \ref{tab:fock_qubit_conversion} (for additional details, and the mapping for the $N=2$ qubit systems, see Supplementary Note 1).

\begin{figure*}
 \centering
 \includegraphics[width=17cm]{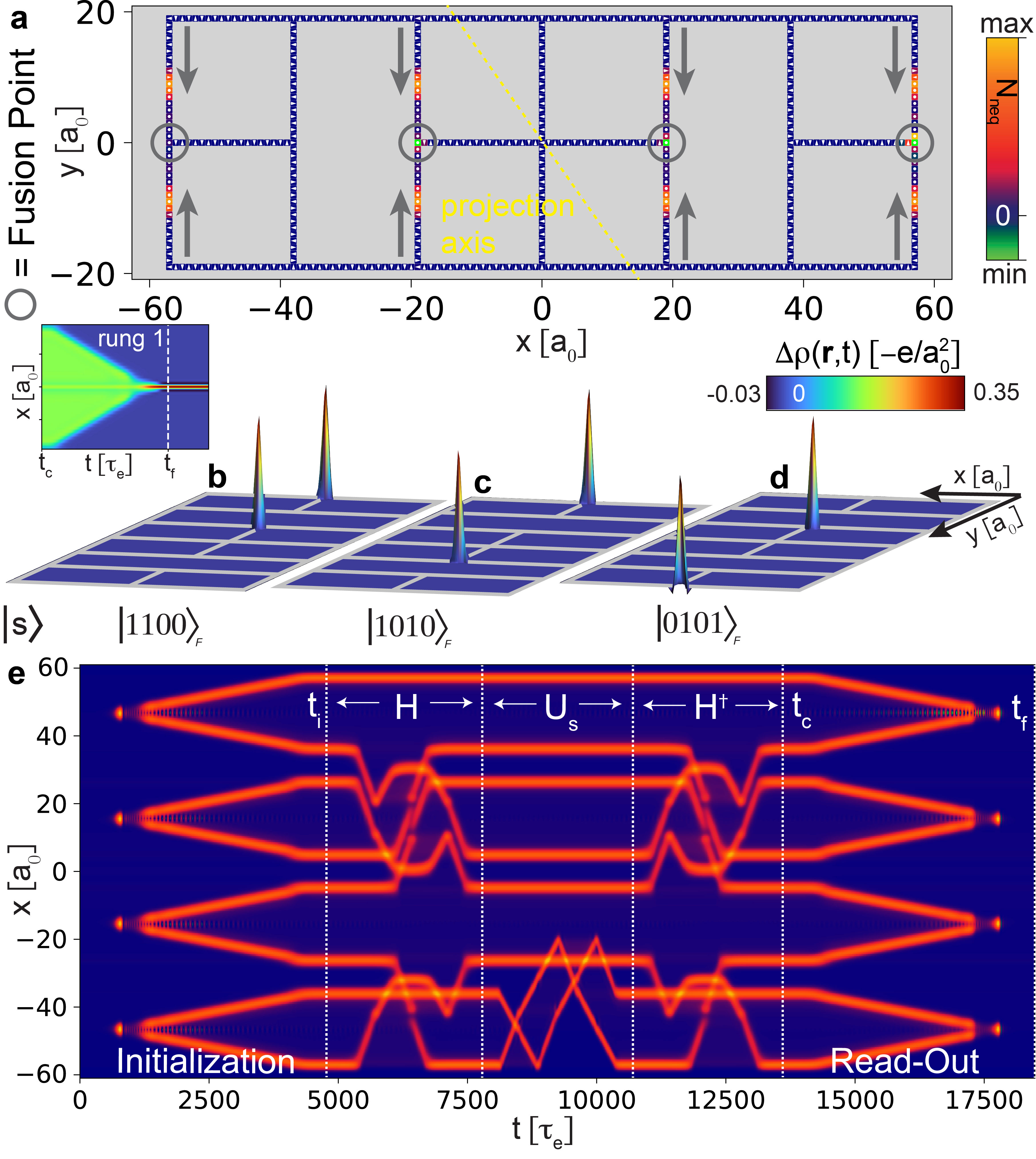}
 \caption{\textbf{Read-out of the final state.}
 {\bf a} Spatial plot of $N_{neq}$ during the fusion process for the $N=3$ qubit system. {\bf b-d} Spatial form of the charge density difference, $\Delta \rho({\bf r})$, after the completion of the fusion process for {\bf b} $s=001$, {\bf c} $s=010$ and {\bf d} $s=100$. Inset of {\bf b} shows the time dependence of $\Delta \rho({\bf r})$ along rung 1. {\bf e} Majorana world lines for the $N=3$ qubit system from initialization to readout, with the real space projection axis shown as a dashed yellow line in {\bf a}. Vertical dotted lines indicate the start and end times for the Hadamard gates and the oracle. Parameters are $(\mu, \alpha, \Delta, JS) = (-3.93, 0.45, 1.2, 2.6) t_e$, $\Gamma = 0.03 t_e$ with execution times of $3200 \tau_e$ for the Hadamard gate, $2460 \tau_e$ for the oracle and $4800 \tau_e$ for the initialization and fusion.}
 \label{fig:Fig3}
\end{figure*}
To test the braiding protocol in Fig.~\ref{fig:Fig2} for the simulation of the BV algorithm, we consider two cases: (i) a $N=2$ qubit system, initialized into the odd-parity state $\ket{n}=\ket{111}_F=\ket{11}_q$ and $s=11$, and (ii) a $N=3$ qubit system initialized into the even-parity state $\ket{n}=\ket{0110}_F=\ket{011}_q$ and $s=001$; the corresponding time-dependent probabilities $p_n(t)$ and $p_{n \oplus s}(t)$ for these two cases are shown in Figs.~\ref{fig:Fig2}{\bf f} and {\bf g}, respectively. Starting from $\ket{GS}$, we obtain for both cases $p_n(t_i) = 1$ (within numerical uncertainty), implying that the system is initialized into the desired state at $t=t_i$. The two peaks in the probabilities (see black arrows) denote the times right after the first Hadamard gate is completed (and the superposition of all $N$-qubit states is achieved), and right before the second Hadamard gate begins. The expected value for the probabilities at these times is $|\bra{n} H \ket{n}|^2 = 1/2^N$ (see horizontal dashed line), in good agreement with our results. After the completion of the BV algorithm at $t=t_c$,
we obtain $p_{n \oplus s}(t_c)=0.997$ and $p_{n \oplus s}(t_c)=0.998$ for the cases shown in Fig.~\ref{fig:Fig2}{\bf f} and {\bf g}, respectively. Since these probabilities, as well as those for all other values of $s$ (see Supplementary Note 2) are below the error correction threshold \cite{Raussendorf_2007,Stace_2009}, our results thus establish the first successful simulation of the BV algorithm in MSH systems.\\

{\it Fusion and read-out.~}
The experimental read-out of the final state $\ket{n \oplus s}$ can be achieved by measuring the charge density arising from the fusion of MZM pairs, as it directly reflects their occupation based on the fusion rule $\gamma \times \gamma = 1 + \psi$, where $\gamma$ represents the MZMs, $1$ the vacuum, and $\psi$ the electron \cite{Beenakker_2019}. After the completion of the BV algorithm, fusion is achieved by rotating the spins in the topological rungs to an AFM in-plane alignment, starting from the ends of the topological rungs and continuing towards their centers, as schematically shown in Fig.~\ref{fig:Fig3}{\bf a}.
In Figs.~\ref{fig:Fig3}{\bf b}-{\bf d}, we present spatial plots of the charge density difference, $\Delta \rho({\bf r},t_f)=\rho({\bf r},t_f)-\rho_{GS}({\bf r})$, after the completion of the fusion process at $t=t_f$. By subtracting the charge density $\rho_{GS}$ of the ground state $\ket{GS}$, we can immediately read out the hidden number $\ket{s}$ in the Fock basis (and then convert it to the logical basis using Table I), by assigning a value of 1 (0) to each MZM pair whose fusion results in a non-zero (zero) residual charge. For example, the BV algorithm for $s=001$ results in a charge density after fusion (see Fig.~\ref{fig:Fig3}{\bf b}) that is nonzero on rungs 1 and 2, and vanishing on rungs 3 and 4, thus yielding $\ket{s}=\ket{1100}_F=\ket{001}_q$ (see Table I), as expected. Similarly, we can identify $\ket{s}$ from the charge density on each rung for $s=010$ (Figs.~\ref{fig:Fig3}{\bf c}) and $s=100$ (Figs.~\ref{fig:Fig3}{\bf d}). A plot of the time-dependent charge density difference along rung 1 is shown in the inset of Figs.~\ref{fig:Fig3}{\bf b}; it demonstrates that the charge remains localized after the completion of the fusion process, as the fusion point is also an intersection point of three rungs which traps the charge and thus facilitates its experimental detection (in contrast to MZM fusion along a rung with no intersection point for the $N=2$ qubit system, see Fig.~\ref{fig:Fig2}{\bf d} and Supplementary Note 3). We thus demonstrated that the hidden number $s$ can be read-out experimentally by measuring the charge density after fusion.\\

{\it Spatial Imaging of Majorana world lines.~}
We visualize the entire BV algorithm from initialization to fusion via the energy-, time- and spatially-resolved non-equilibrium density of states $N_{neq}({\bf r}, \sigma, t, \omega)$ \cite{Bedow_2022} (for details, see Methods Section). The full time dependence of the spatially resolved zero-energy $N_{neq}$ for the cases of Figs.~\ref{fig:Fig2}{\bf f} and {\bf g} is provided in Supplementary Movies 1 and 2, respectively (see also Supplementary Figs.~3 and 4 in Supplementary Note 4), demonstrating, as expected, that at any time, the MZMs are localized near the edges of the topological regions. By projecting $N_{neq}$ onto the real space axis shown in Fig.~\ref{fig:Fig3}\textbf{a}, we image the Majorana world lines visualizing the entire BV algorithm in time and space, as shown in Fig.~\ref{fig:Fig3}{\bf e} (a full 3D rendering of the world lines for the cases of Figs.~\ref{fig:Fig2}{\bf f} and {\bf g} is shown in Supplementary Movies 3 and 4, respectively).\\

\begin{figure*}
 \centering
 \includegraphics[width=17cm]{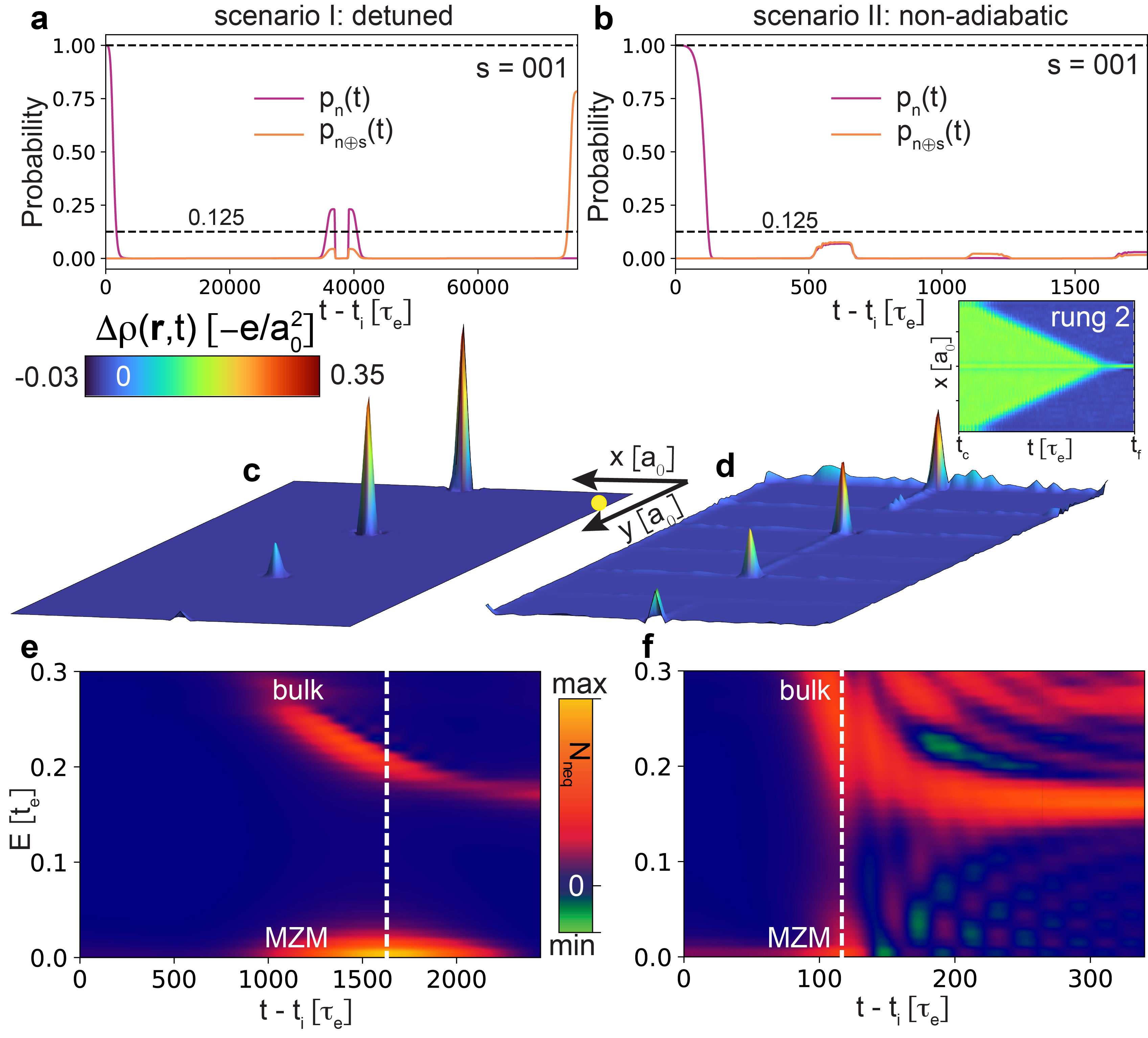}
 \caption{\textbf{Faulty Implementations of the BV algorithm.}
 Time-dependent transition probabilities for s=111 and {\bf a} a detuned Hadamard gate (scenario I), and
 {\bf b} a non-adiabatic implementation with a significantly reduced execution time of the BV algorithm (scenario II). $\Delta \rho({\bf r})$ after completion of the fusion process for $s=100$ and {\bf c} scenario I, and {\bf d} scenario II. Energy-resolved non-equilibrium local density of states at the site marked by the yellow dot in {\bf c} for {\bf e} scenario I and {\bf f} scenario II, showing the movement of one of the Majorana zero modes through this site.
 Parameters are $(\mu, \alpha, \Delta, JS) = (-3.93, 0.45, 1.2, 2.6) t_e$, $\Gamma = 0.03 t_e$ and execution times of $36800 \tau_e$ ($640.0 \tau_e$) for the Hadamard gate, $2460 \tau_e$ ($496 \tau_e$) for the oracle, and $4800 \tau_e$($960 \tau_e$) for the initialization and fusion processes for scenario I (II).}
 \label{fig:Fig4}
\end{figure*}

{\it Detecting faulty implementations.~}
The question naturally arises to what extent faulty implementations of the BV algorithm with $p_{n \oplus s}(t_c)<1$ can be experimentally detected.
To investigate this question, we consider two faulty scenarios: (I) we detune the Hadamard gate, by changing its execution time such that, while the BV algorithm is still adiabatic, the state created after the application of the Hadamard gate is not an equal superposition of pure qubit states (see Supplementary Note 1), and (II) by significantly reducing the execution time of the entire BV algorithm by a factor of 5, which leads to excitations between the MZMs and bulk states and thus renders the BV algorithm non-adiabatic. The time-dependent probabilities for both scenarios with $s=001$ are shown in Figs.~\ref{fig:Fig4}{\bf a} and {\bf b}, respectively, revealing significantly reduced success probability of $p_{n \oplus s}(t_c)=0.78$ (scenario I) and $p_{n \oplus s}(t_c)=0.0167$ (scenario II).
 These are directly reflected in the spatial form of $\Delta \rho$: while for the perfect, adiabatic BV algorithm for $s=001$, a non-zero charge distribution after fusion exists only on rungs 1 and 2
(see Fig.~\ref{fig:Fig3}{\bf b}), the imperfect scenarios I and II now also yield non-zero charges on rungs 3 and 4 (see Figs.~\ref{fig:Fig4}{\bf c} and {\bf d}, respectively). In addition, for scenario II, a non-zero $\Delta \rho({\bf r})$ emerges even away from the fusion point, thus discriminating it not only form the perfect case, but also from scenario I. Moreover, scenarios I and II can also be distinguished using the time- and energy-dependent $N_{neq}$ (see Figs.~\ref{fig:Fig4}{\bf e} and {\bf f}), at a specific site in the MSH network indicated by the yellow dot in Figs.~\ref{fig:Fig4}{\bf c}. In both cases, the MZM moves through this site during the BV algorithm, with the maximum intensity indicated by a dashed white line. However, for the adiabatic scenario I, the MZM remains well separated from the bulk states, while for the non-adiabatic scenario II, a hybridization between the MZM and the bulk states is clearly visible (see white dashed line in Fig.~\ref{fig:Fig4}{\bf f}). We thus conclude that the form of $\Delta \rho$ as well as that of $N_{neq}$ distinguishes not only the faulty implementation of the BV algorithm from a perfect one, but also identify the origin of the faulty realization, -- detuning versus non-adiabaticity.\\

{\it Discussion.~}
We have demonstrated the successful simulation of the topologically protected BV algorithm from initialization to read-out in 2D MSH systems using Majorana zero modes. We showed that starting from the ground state of the topologically trivial systems, it is possible to initialize the system into various topological even- and odd-parity states through the interplay of the MSH architecture and the system's chemical potential. We demonstrated how the hidden number $s$ in the logical qubit bases can be straightforwardly encoded in the braiding protocol representing the oracle, allowing us to arrive at an efficient translation between the Majorana-based Fock notation and the logical qubit notation for the many-body states. We proposed a new approach to efficiently and uniquely generalize the Hadamard gate to $N \geq 2$ qubit systems using simultaneously executed $\sqrt{X}$-gates, thus ensuring that the execution time of the Hadamard gate is independent of $N$. We also showed that after the conclusion of the BV algorithm and the fusion of the MZM pairs, the hidden number $s$ can be read out from the spatial form of the excess charge density, $\Delta \rho$.
We demonstrated that the entire BV algorithm is visualized in time and space via the non-equilibrium density of states, $N_{neq}$, which allowed us to directly image the Majorana world lines. Finally, we showed that the form of $\Delta \rho$ and $N_{neq}$ allows us to discriminate between perfect and faulty implementations of the BV algorithm, and identify the origin of the latter. Our results thus provide the proof of concept that topologically protected, complex quantum algorithms can be successfully simulated in MSH structures.\\

While experimental challenges and approaches to the implementation of the above braiding scheme have been discussed before \cite{Bedow2024}, several theoretical questions remain open and need to be addressed in the future. Of particular importance are here the effects of decoherence, noise and quasi-particle poisoning -- i.e., the presence of unpaired electrons -- on the successful simulation of the algorithm. Our preliminary work on single $X$- and $Z$-gates has shown that in MSH systems, the effect of quasi-particle poisoning is negligible, if the braiding process is conducted in the adiabatic limit. A further discussion of this and other pertinent issues is reserved for future work.\\

{\bf Methods} \\
{\it Creation of the Topological Computation Basis} \\
The Hamiltonian in Eq.~\eqref{eq:H} can be rewritten in the Bogoliubov-de Gennes (BdG) form
\begin{equation}
    \mathcal{H} = \frac{1}{2} \sum_{i,j}
    \begin{pmatrix}
        c_i^\dagger & c_i
    \end{pmatrix}
    \underbrace{
    \begin{pmatrix}
        H_{ij} (t) & \Delta_{ij} \\
        \Delta^{*}_{ji} & -H^{*}_{ij} (t)
    \end{pmatrix}}_{=H_\mathrm{BdG}}
    \begin{pmatrix}
        c_j \\
        c_j^\dagger
    \end{pmatrix} \; ,
\end{equation}
where $H_\mathrm{BdG}$ possesses a particle-hole symmetry, which can be expressed as $H_\mathrm{BdG} = - \tau_x H^{*}_\mathrm{BdG} \tau_x$ with $\tau_x$ being the Pauli X-matrix in particle-hole space.
At $t=0$, the Bogoliubov transformation
\begin{equation}
    \begin{pmatrix}
        c_j \\
        c_j^\dagger
    \end{pmatrix}
    =
    \sum_{n}
    \begin{pmatrix}
        U_{jn} & V_{jn}^{*} \\
        V_{jn} & U_{jn}^{*}
    \end{pmatrix}
    \begin{pmatrix}
        d_n \\
        d_n^\dagger
    \end{pmatrix}
    \label{eq:Bdg}
\end{equation}
diagonalizes the Hamiltonian
\begin{equation}
    \mathcal{H} = \sum_{n} E_n (d_n^\dagger d_n - \frac{1}{2}) \; ,
\end{equation}
with $E_n \geq 0$. The ground state $\ket{\Omega}$ of the system is the quasiparticle vacuum defined via $d_n \ket{\Omega} = 0 \; \forall \; n$.  We construct it from the vacuum of the $c$-operators $\ket{0}$ \cite{Ring_1980,Alicea_2011} via
%by annihilating all $d$-operators
\begin{equation}
    \ket{\Omega} = \frac{1}{\sqrt{\mathcal{N}}} \prod_n d_n \ket{0} \; ,
\end{equation}
where the normalization is given by $\mathcal{N} = |\mathrm{det} (\hat{V} )|$, where $[\hat{V}]_{jn} = V_{jn}$ in Eq.~(\ref{eq:Bdg}).
As the Hamiltonian considered here conserves fermion parity, we restrict ourselves to either the even or the odd parity sector, i.e. $P = \sum_n \langle d_n^\dagger d_n \rangle = 0 \; \text{or} \; 1 \; \text{mod} \; 2$.
The states in our topological computation basis with $N$ qubits, i.e. $N+1$ Majorana pairs are thus
\begin{align}
    \ket{n_1 \cdots n_N n_N+1}_F = \prod_{j=1}^N \left(d_j^\dagger\right)^{n_j} \left(d_{N+1}^\dagger\right)^{\bigoplus_{j=1}^N n_j \oplus P} \ket{\Omega}.
\end{align}
Here, $d_j^\dagger$ creates an electron in the $j^{th}$ Majorana pair (i.e., this pair becomes occupied), and the occupation of the $(N+1)$-st Majorana pair is chosen such that the fermion parity $P$ of all $2^N$ states is fixed to be either even ($P=0$) or odd ($P=1$). Which one of the Majorana pairs is chosen as the parity-conserving one is arbitrary.\\

{\it Time Evolution of States} \\
The time-evolved quasiparticle operators are given by
\begin{equation}
    d_n(t) = \mathcal{U}(t) d_n \mathcal{U}^\dagger(t)
\end{equation}
with the time-evolution operator
\begin{equation}
    \mathcal{U}(t) = \mathcal{T} \exp\left( - \frac{\mathrm{i}}{\hbar} \int_0^t \mathrm{d}t' \mathcal{H}(t') \right) \; .
\end{equation}
Using the time-dependent BdG equations, the time-evolved quasiparticle operators are given by
\begin{equation}
%    \begin{pmatrix} d_n^\dagger(t) & d_n(t) \end{pmatrix}
        \begin{pmatrix} d_n^\dagger(t), d_n(t) \end{pmatrix}
    = \sum_i
    \begin{pmatrix} c_i^\dag & c_i \end{pmatrix}
    \begin{pmatrix}
        U_{in}(t) & V^*_{in}(t) \\[10pt]
        V_{in}(t) & U^*_{in}(t)
    \end{pmatrix},
    \label{eq:Bdg_timedep}
\end{equation}
where
\begin{widetext}
\begin{align}
    \begin{pmatrix}
        \hat{U}(t) & \hat{V}^*(t) \\[10pt]
        \hat{V}(t) & \hat{U}^*(t)
    \end{pmatrix}
    &= \underbrace{\mathcal{T}\exp\left[
        -\frac{\mathrm{i}}{\hbar} \int_0^t dt' H_{\text{BdG}}(t')
    \right]}_{=U_\mathrm{BdG}}
    \begin{pmatrix}
        \hat{U}(0) & \hat{V}(0)^* \\[10pt]
        \hat{V}(0) & \hat{U}^*(0)
    \end{pmatrix} \; ,
\end{align}
\end{widetext}
with the elements of the matrices $\hat{U}, \hat{V}$ given by $[\hat{U}]_{in} = U_{in}$ and  $[\hat{V}]_{in} = V_{in}$ in Eq.~(\ref{eq:Bdg_timedep}).
As such, we can write the time-evolved quasiparticle vacuum as
\begin{align}
    \ket{\Omega(t)} = \frac{e^{\mathrm{i}\alpha(t)}}{\sqrt{\mathcal{N}(t)}}
    \prod_{k} d_k(t) \ket{0} \ ,
\end{align}
with the normalization $\mathcal{N}(t) = |\det(V(t))|$. The phase $\alpha$ stems from the time evolution of the true vacuum.
However, this phase is gauged away in our gauge-invariant formulation of physical quantities, such as the transition probabilities.
Hence, the time-evolved states in our logical basis are given by
\begin{equation}
\begin{aligned}
    &\ket{n_1 \cdots n_N n_N+1}_F (t) \\
    &\; = \prod_{j=1}^N \left(d_j^\dagger (t)\right)^{n_j} \left(d_{N+1}^\dagger (t)\right)^{\bigoplus_{j=1}^N n_j \oplus P} \ket{\Omega (t)} \ .
\end{aligned}
\end{equation}

{\it Transition Probabilities} \\
The probability for transitions between states $\ket{\psi(t)} = \ket{n_1 \cdots n_N n_{N+1}}_F (t)$ and $\ket{\psi' (0)} = \ket{n_1' \cdots n_N' n_{N+1}'}_F (0)$ takes the form
\begin{widetext}
\begin{align}
    |\braket{\psi'(0) | \psi(t)}|^2
    = \left|(-1)^s \frac{e^{\mathrm{i}\alpha(t)}}{\sqrt{\mathcal{N} \mathcal{N}(t)}}
    \braket{0|
        \prod_{l} d_l^\dagger
        \prod_{j=1}^N \left(d_j \right)^{n'_j} \left(d_{N+1} \right)^{\bigoplus_{j=1}^N n'_j \oplus P}
        \prod_{j=1}^N \left(d_j^\dagger (t)\right)^{n_j} \left(d_{N+1}^\dagger (t)\right)^{\bigoplus_{j=1}^N n_j \oplus P}
        \prod_{l} d_l(t)
    |0} \right|^2.
\end{align}
\end{widetext}
This probability can be computed using the Onishi Formula \cite{Onishi_1966, Ring_1980} and the Bogoliubov transformations from Eqs.~\eqref{eq:Bdg} and ~\eqref{eq:Bdg_timedep} via
\begin{equation}
    \left|\braket{\psi'(0) | \psi(t)}\right|^2 = \left|\mathrm{det} [\hat{U}'^\dagger \hat{U} (t) + \hat{V}'^\dagger \hat{V} (t)]\right| \; .
\end{equation}

{\it Charge Density} \\
The electronic charge density at site ${\bf r}$ for spin $\sigma$  in the state $\ket{\psi}$ is given by $\rho({\bf r}, \sigma) = -e n_{{\bf r},\sigma}$, where
\begin{equation}
    n_{{\bf r},\sigma} = \braket{\psi|c^\dagger_{{\bf r}, \sigma} c_{{\bf r}, \sigma}|\psi} \; .
\end{equation}
Numbering the sites and other degrees of freedom from 1 to $N_\mathrm{D}$, we define the spinors
\begin{align}
    \Psi &= \begin{pmatrix}
        c_{1} \\
        \vdots \\
        c_{N_\mathrm{D}} \\
        c^\dagger_{1} \\
        \vdots \\
        c^\dagger_{N_\mathrm{D}}
    \end{pmatrix}
    \quad\text{and}\quad
    \Phi = \begin{pmatrix}
        d_{1} \\
        \vdots \\
        d_{N_\mathrm{D}} \\
        d^\dagger_{1} \\
        \vdots \\
        d^\dagger_{N_\mathrm{D}}
    \end{pmatrix} \;, 
\end{align}
which are related via the Bogulubov-de Gennes transformation as 
\begin{equation}
    \Psi = M \Phi \; \text{with} \; M = \begin{pmatrix}
        \hat{U} & \hat{V}^{*} \\
        \hat{V} & \hat{U}^{*}
    \end{pmatrix}
\end{equation}
where the elements of the matrices $\hat{U}$, $\hat{V}$ given by $[\hat{U}]_{in} = U_{in}$ and  $[\hat{V}]_{in} = V_{in}$ in Eq.~(\ref{eq:Bdg_timedep}).
We have for the ground state $\ket{\Omega}$
\begin{align}
    \braket{\Omega|\Phi \Phi^\dagger|\Omega} &= 
    \begin{pmatrix}
        I_{N\times N} & 0 \\
        0 & 0 \\
    \end{pmatrix} \ ,
\end{align}
and thus
\begin{equation}
\begin{aligned}
    \braket{\Omega|\Psi \Psi^\dagger|\Omega} &= \braket{\Omega|M \Phi \Phi^\dagger M^\dagger |\Omega} = M \braket{\Omega|\Phi \Phi^\dagger|\Omega} M^\dagger \\
    &= \begin{pmatrix}
        \hat{U} & \hat{V}^{*} \\
        \hat{V} & \hat{U}^{*}
    \end{pmatrix} \begin{pmatrix}
        I_{N_\mathrm{DoF}\times N_\mathrm{DoF}} & 0 \\
        0 & 0 \\
    \end{pmatrix}
    \begin{pmatrix}
        \hat{U}^\dagger & \hat{V}^{\dagger} \\
        \hat{V}^{T} & \hat{U}^{T}
    \end{pmatrix} \\
    &= \begin{pmatrix}
        \hat{U} \hat{U}^\dagger & \hat{U} \hat{V}^\dagger \\
        \hat{V} \hat{U}^\dagger & \hat{V} \hat{V}^\dagger 
    \end{pmatrix}  \ .
\end{aligned}
\end{equation}
The charge density is then given by the $N_\mathrm{D}$ diagonal elements of the lower right block of this matrix and thus
\begin{equation}
    n_i = \left(\hat{V} \hat{V}^\dagger \right)_{ii} \; .
\end{equation}
The time-evolved charge density is obtained by using the time-evolved Boguliubov-de Gennes equations, and the resulting time-evolved matrix $M(t)$
\begin{align}
    \Psi(t) = \mathcal{U}(t) \Psi \mathcal{U}^\dagger(t) = 
    U_\text{BdG} \Psi = U_\text{BdG} M \Phi = M(t) \Phi \ .
\end{align}
The time-dependent charge density is then obtained from
\begin{equation}
    n_i(t) = \left(\hat{V}(t) \hat{V}^\dagger(t) \right)_{ii} \; ,
\end{equation}
as 
\begin{equation}
    \rho({\bf r}, \sigma, t) = -e \left(\hat{V}(t) \hat{V}^\dagger(t) \right)_{({\bf r}, \sigma),({\bf r}, \sigma)} \; .
\end{equation}
\phantom{a}\\

{\it Time Dependence of Spin Rotations} \\
The rotation of the spins in the network is governed by two time scales during each step of the algorithm: a rotation time $T_R$ to rotate one individual spin by a polar angle of $\pi/2$, and a delay time $\Delta T_R$ before the next spin starts rotating.
We use spherical coordinates to describe each spin's orientation in space, such that
\begin{align}
{\bf S}_{\bf R}(t)  =  {
\begin{pmatrix}
    \cos(\phi({\bf R},t) \cdot \sin(\theta({\bf R},t)) \\[3pt]
    \sin(\phi({\bf R},t) \cdot \sin(\theta({\bf R},t)) \\[3pt]
    \cos(\theta({\bf R},t)) \\
\end{pmatrix}\ .
}
\end{align}
Therein, we use the function
\begin{equation}
    s(t, t_0) =
    \frac{\pi}{2}
    \begin{cases}
    0 ,& t < t_0 \\
    \sin^2\left( \frac{t-t_0}{T_\text{R}}\right) ,& t_0 \leq t \leq t_0 + T_\text{R} \\
    1 ,&  t > t_0 + T_\text{R}
    \end{cases}
\end{equation}
so that the transition from 0 to $\pi/2$ for the polar angle is executed smoothly.
For instance, the initialization of the spins ${\bf S}_i$ on one rung of length $2L+3$, numbered from 0 to $2L+2$, from in-plane antiferromagnetic alignment to out-of-plane ferromagnetic alignment is executed by the following time dependence of the polar angles
\begin{widetext}
\begin{equation}
    (\theta_i(t)) = \begin{cases}
        (-1)^i \frac{\pi}{2} (1 - s(t, \Delta T_R \cdot (L + 1 - i)), & 0 \leq i \leq L+1 \\
        (-1)^i \frac{\pi}{2} (1 - s(t, \Delta T_R \cdot (i - L - 1)),& L+2 \leq i \leq 2L+2
    \end{cases}
    \; ,
\end{equation}
\end{widetext}
which rotates the center spin out-of-plane first, followed by the next two adjacent spins after the delay time $\Delta T_R$, and so forth until all spins are rotated out-of-plane. The azimuthal angle $\phi$ is chosen for each spin such that its in-plane orientation is perpendicular to the direction of the network chain at that site, i.e. for a horizontal segment $\phi=\frac{\pi}{2}$, for a vertical segment $\phi=0$.
All other parts of the processes are executed with similar rotation protocols.

{\it Non-equilibrium Local Density of States} \\
The time-dependent processes discussed in the main text can be experimentally visualized by measuring the time-dependent differential conductance $dI(V,{\bf r},t)/dV$ in scanning tunneling microscopy (STM) experiments.
We previously showed that, in analogy to the equilibrium case, this quantity is proportional to the non-equilibrium density of states defined as $N_{neq}({\bf r}, \sigma, t, \omega) = -\frac{1}{\pi} {\rm Im} G^{r}({\bf r},{\bf r}, \sigma, t, \omega)$ \cite{Bedow_2022}. Here, the time- and frequency-dependent retarded Greens function $\hat{G}^r$ is calculated from the differential equation
\begin{equation}
    \left[ \mathrm{i} \frac{d}{dt} +  \omega  + \mathrm{i} \Gamma - {\hat H}(t)\right] {\hat G}^r \left(t, \omega \right) = {\hat 1} \; .
\end{equation}

{\it Acknowledgments}
This work was supported by the U.\ S.\ Department of Energy, Office of Science, Basic Energy Sciences, under Award No.\ DE-FG02-05ER46225. We would like to acknowledge helpful discussions with C. Lutz, E. Mascot, S. Rachel, and R. Wiesendanger.

{\bf Data availability}
Original data are available at (insert link to Zenodo depository).\\

{\bf Code availability}
The codes that were employed in this study are available from the authors on reasonable request.\\

{\bf Author contributions}
J.B performed the theoretical calculations. D.K.M. devised the project and supervised the theoretical calculations. D.K.M. and J.B. discussed the results and wrote the manuscript.\\

{\bf Competing interests}
The authors declare no competing interests.

%\bibliographystyle{naturemag}
%\bibliographystyle{sn-mathphys}
%\bibliography{MSH_Quantum_Alg}

\begin{thebibliography}{54}%
    \makeatletter
    \providecommand \@ifxundefined [1]{%
     \@ifx{#1\undefined}
    }%
    \providecommand \@ifnum [1]{%
     \ifnum #1\expandafter \@firstoftwo
     \else \expandafter \@secondoftwo
     \fi
    }%
    \providecommand \@ifx [1]{%
     \ifx #1\expandafter \@firstoftwo
     \else \expandafter \@secondoftwo
     \fi
    }%
    \providecommand \natexlab [1]{#1}%
    \providecommand \enquote  [1]{``#1''}%
    \providecommand \bibnamefont  [1]{#1}%
    \providecommand \bibfnamefont [1]{#1}%
    \providecommand \citenamefont [1]{#1}%
    \providecommand \href@noop [0]{\@secondoftwo}%
    \providecommand \href [0]{\begingroup \@sanitize@url \@href}%
    \providecommand \@href[1]{\@@startlink{#1}\@@href}%
    \providecommand \@@href[1]{\endgroup#1\@@endlink}%
    \providecommand \@sanitize@url [0]{\catcode `\\12\catcode `\$12\catcode
      `\&12\catcode `\#12\catcode `\^12\catcode `\_12\catcode `\%12\relax}%
    \providecommand \@@startlink[1]{}%
    \providecommand \@@endlink[0]{}%
    \providecommand \url  [0]{\begingroup\@sanitize@url \@url }%
    \providecommand \@url [1]{\endgroup\@href {#1}{\urlprefix }}%
    \providecommand \urlprefix  [0]{URL }%
    \providecommand \Eprint [0]{\href }%
    \providecommand \doibase [0]{https://doi.org/}%
    \providecommand \selectlanguage [0]{\@gobble}%
    \providecommand \bibinfo  [0]{\@secondoftwo}%
    \providecommand \bibfield  [0]{\@secondoftwo}%
    \providecommand \translation [1]{[#1]}%
    \providecommand \BibitemOpen [0]{}%
    \providecommand \bibitemStop [0]{}%
    \providecommand \bibitemNoStop [0]{.\EOS\space}%
    \providecommand \EOS [0]{\spacefactor3000\relax}%
    \providecommand \BibitemShut  [1]{\csname bibitem#1\endcsname}%
    \let\auto@bib@innerbib\@empty
    %</preamble>
    \bibitem [{\citenamefont {Nayak}\ \emph {et~al.}(2008)\citenamefont {Nayak},
      \citenamefont {Simon}, \citenamefont {Stern}, \citenamefont {Freedman},\ and\
      \citenamefont {Das~Sarma}}]{Nayak_2008}%
      \BibitemOpen
      \bibfield  {author} {\bibinfo {author} {\bibfnamefont {C.}~\bibnamefont
      {Nayak}}, \bibinfo {author} {\bibfnamefont {S.~H.}\ \bibnamefont {Simon}},
      \bibinfo {author} {\bibfnamefont {A.}~\bibnamefont {Stern}}, \bibinfo
      {author} {\bibfnamefont {M.}~\bibnamefont {Freedman}},\ and\ \bibinfo
      {author} {\bibfnamefont {S.}~\bibnamefont {Das~Sarma}},\ }\bibfield  {title}
      {\bibinfo {title} {Non-{{Abelian}} anyons and topological quantum
      computation},\ }\href {https://doi.org/10.1103/RevModPhys.80.1083} {\bibfield
       {journal} {\bibinfo  {journal} {Rev. Mod. Phys.}\ }\textbf {\bibinfo
      {volume} {80}},\ \bibinfo {pages} {1083} (\bibinfo {year}
      {2008})}\BibitemShut {NoStop}%
    \bibitem [{\citenamefont {Sarma}\ \emph {et~al.}(2015)\citenamefont {Sarma},
      \citenamefont {Freedman},\ and\ \citenamefont {Nayak}}]{Sarma_2015}%
      \BibitemOpen
      \bibfield  {author} {\bibinfo {author} {\bibfnamefont {S.~D.}\ \bibnamefont
      {Sarma}}, \bibinfo {author} {\bibfnamefont {M.~H.}\ \bibnamefont
      {Freedman}},\ and\ \bibinfo {author} {\bibfnamefont {C.}~\bibnamefont
      {Nayak}},\ }\bibfield  {title} {\bibinfo {title} {Majorana zero modes and
      topological quantum computation},\ }\href
      {https://doi.org/10.1038/npjqi.2015.1} {\bibfield  {journal} {\bibinfo
      {journal} {npj Quantum Inf.}\ }\textbf {\bibinfo {volume} {1}},\ \bibinfo
      {pages} {15001} (\bibinfo {year} {2015})}\BibitemShut {NoStop}%
    \bibitem [{\citenamefont {Beenakker}(2020)}]{Beenakker_2019}%
      \BibitemOpen
      \bibfield  {author} {\bibinfo {author} {\bibfnamefont {C.~W.~J.}\
      \bibnamefont {Beenakker}},\ }\bibfield  {title} {\bibinfo {title} {{Search
      for non-Abelian Majorana braiding statistics in superconductors}},\ }\href
      {https://doi.org/10.21468/SciPostPhysLectNotes.15} {\bibfield  {journal}
      {\bibinfo  {journal} {SciPost Phys. Lect. Notes}\ ,\ \bibinfo {pages} {15}}
      (\bibinfo {year} {2020})}\BibitemShut {NoStop}%
    \bibitem [{\citenamefont {Nadj-Perge}\ \emph {et~al.}(2014)\citenamefont
      {Nadj-Perge}, \citenamefont {Drozdov}, \citenamefont {Li}, \citenamefont
      {Chen}, \citenamefont {Jeon}, \citenamefont {Seo}, \citenamefont {MacDonald},
      \citenamefont {Bernevig},\ and\ \citenamefont {Yazdani}}]{Nadj-Perge_2014}%
      \BibitemOpen
      \bibfield  {author} {\bibinfo {author} {\bibfnamefont {S.}~\bibnamefont
      {Nadj-Perge}}, \bibinfo {author} {\bibfnamefont {I.~K.}\ \bibnamefont
      {Drozdov}}, \bibinfo {author} {\bibfnamefont {J.}~\bibnamefont {Li}},
      \bibinfo {author} {\bibfnamefont {H.}~\bibnamefont {Chen}}, \bibinfo {author}
      {\bibfnamefont {S.}~\bibnamefont {Jeon}}, \bibinfo {author} {\bibfnamefont
      {J.}~\bibnamefont {Seo}}, \bibinfo {author} {\bibfnamefont {A.~H.}\
      \bibnamefont {MacDonald}}, \bibinfo {author} {\bibfnamefont {B.~A.}\
      \bibnamefont {Bernevig}},\ and\ \bibinfo {author} {\bibfnamefont
      {A.}~\bibnamefont {Yazdani}},\ }\bibfield  {title} {\bibinfo {title}
      {Observation of majorana fermions in ferromagnetic atomic chains on a
      superconductor},\ }\href {https://doi.org/10.1126/science.1259327} {\bibfield
       {journal} {\bibinfo  {journal} {Science}\ }\textbf {\bibinfo {volume}
      {346}},\ \bibinfo {pages} {602} (\bibinfo {year} {2014})}\BibitemShut
      {NoStop}%
    \bibitem [{\citenamefont {Ruby}\ \emph {et~al.}(2015)\citenamefont {Ruby},
      \citenamefont {Pientka}, \citenamefont {Peng}, \citenamefont {von Oppen},
      \citenamefont {Heinrich},\ and\ \citenamefont {Franke}}]{Ruby_2015}%
      \BibitemOpen
      \bibfield  {author} {\bibinfo {author} {\bibfnamefont {M.}~\bibnamefont
      {Ruby}}, \bibinfo {author} {\bibfnamefont {F.}~\bibnamefont {Pientka}},
      \bibinfo {author} {\bibfnamefont {Y.}~\bibnamefont {Peng}}, \bibinfo {author}
      {\bibfnamefont {F.}~\bibnamefont {von Oppen}}, \bibinfo {author}
      {\bibfnamefont {B.~W.}\ \bibnamefont {Heinrich}},\ and\ \bibinfo {author}
      {\bibfnamefont {K.~J.}\ \bibnamefont {Franke}},\ }\bibfield  {title}
      {\bibinfo {title} {End states and subgap structure in proximity-coupled
      chains of magnetic adatoms},\ }\href
      {https://doi.org/10.1103/PhysRevLett.115.197204} {\bibfield  {journal}
      {\bibinfo  {journal} {Phys. Rev. Lett.}\ }\textbf {\bibinfo {volume} {115}},\
      \bibinfo {pages} {197204} (\bibinfo {year} {2015})}\BibitemShut {NoStop}%
    \bibitem [{\citenamefont {Pawlak}\ \emph {et~al.}(2016)\citenamefont {Pawlak},
      \citenamefont {Kisiel}, \citenamefont {Klinovaja}, \citenamefont {Meier},
      \citenamefont {Kawai}, \citenamefont {Glatzel}, \citenamefont {Loss},\ and\
      \citenamefont {Meyer}}]{Pawlak_2016}%
      \BibitemOpen
      \bibfield  {author} {\bibinfo {author} {\bibfnamefont {R.}~\bibnamefont
      {Pawlak}}, \bibinfo {author} {\bibfnamefont {M.}~\bibnamefont {Kisiel}},
      \bibinfo {author} {\bibfnamefont {J.}~\bibnamefont {Klinovaja}}, \bibinfo
      {author} {\bibfnamefont {T.}~\bibnamefont {Meier}}, \bibinfo {author}
      {\bibfnamefont {S.}~\bibnamefont {Kawai}}, \bibinfo {author} {\bibfnamefont
      {T.}~\bibnamefont {Glatzel}}, \bibinfo {author} {\bibfnamefont
      {D.}~\bibnamefont {Loss}},\ and\ \bibinfo {author} {\bibfnamefont
      {E.}~\bibnamefont {Meyer}},\ }\bibfield  {title} {\bibinfo {title} {Probing
      atomic structure and majorana wavefunctions in mono-atomic fe-chains on
      superconducting pb-surface},\ }\href {https://doi.org/10.1038/npjqi.2016.35}
      {\bibfield  {journal} {\bibinfo  {journal} {npj Quantum Inf.}\ }\textbf
      {\bibinfo {volume} {2}},\ \bibinfo {pages} {16035} (\bibinfo {year}
      {2016})}\BibitemShut {NoStop}%
    \bibitem [{\citenamefont {Kim}\ \emph {et~al.}(2018)\citenamefont {Kim},
      \citenamefont {Palacio-Morales}, \citenamefont {Posske}, \citenamefont
      {Rózsa}, \citenamefont {Palotás}, \citenamefont {Szunyogh}, \citenamefont
      {Thorwart},\ and\ \citenamefont {Wiesendanger}}]{Kim_2018}%
      \BibitemOpen
      \bibfield  {author} {\bibinfo {author} {\bibfnamefont {H.}~\bibnamefont
      {Kim}}, \bibinfo {author} {\bibfnamefont {A.}~\bibnamefont
      {Palacio-Morales}}, \bibinfo {author} {\bibfnamefont {T.}~\bibnamefont
      {Posske}}, \bibinfo {author} {\bibfnamefont {L.}~\bibnamefont {Rózsa}},
      \bibinfo {author} {\bibfnamefont {K.}~\bibnamefont {Palotás}}, \bibinfo
      {author} {\bibfnamefont {L.}~\bibnamefont {Szunyogh}}, \bibinfo {author}
      {\bibfnamefont {M.}~\bibnamefont {Thorwart}},\ and\ \bibinfo {author}
      {\bibfnamefont {R.}~\bibnamefont {Wiesendanger}},\ }\bibfield  {title}
      {\bibinfo {title} {Toward tailoring majorana bound states in artificially
      constructed magnetic atom chains on elemental superconductors},\ }\href
      {https://doi.org/10.1126/sciadv.aar5251} {\bibfield  {journal} {\bibinfo
      {journal} {Sci. Adv.}\ }\textbf {\bibinfo {volume} {4}},\ \bibinfo {pages}
      {eaar5251} (\bibinfo {year} {2018})}\BibitemShut {NoStop}%
    \bibitem [{\citenamefont {Palacio-Morales}\ \emph {et~al.}(2019)\citenamefont
      {Palacio-Morales}, \citenamefont {Mascot}, \citenamefont {Cocklin},
      \citenamefont {Kim}, \citenamefont {Rachel}, \citenamefont {Morr},\ and\
      \citenamefont {Wiesendanger}}]{Palacio-Morales2019}%
      \BibitemOpen
      \bibfield  {author} {\bibinfo {author} {\bibfnamefont {A.}~\bibnamefont
      {Palacio-Morales}}, \bibinfo {author} {\bibfnamefont {E.}~\bibnamefont
      {Mascot}}, \bibinfo {author} {\bibfnamefont {S.}~\bibnamefont {Cocklin}},
      \bibinfo {author} {\bibfnamefont {H.}~\bibnamefont {Kim}}, \bibinfo {author}
      {\bibfnamefont {S.}~\bibnamefont {Rachel}}, \bibinfo {author} {\bibfnamefont
      {D.~K.}\ \bibnamefont {Morr}},\ and\ \bibinfo {author} {\bibfnamefont
      {R.}~\bibnamefont {Wiesendanger}},\ }\bibfield  {title} {\bibinfo {title}
      {{Atomic-scale interface engineering of Majorana edge modes in a 2D
      magnet-superconductor hybrid system}},\ }\href
      {https://doi.org/10.1126/sciadv.aav6600} {\bibfield  {journal} {\bibinfo
      {journal} {Science Advances}\ }\textbf {\bibinfo {volume} {5}} (\bibinfo
      {year} {2019})}\BibitemShut {NoStop}%
    \bibitem [{\citenamefont {M{\'e}nard}\ \emph {et~al.}(2017)\citenamefont
      {M{\'e}nard}, \citenamefont {Guissart}, \citenamefont {Brun}, \citenamefont
      {Leriche}, \citenamefont {Trif}, \citenamefont {Debontridder}, \citenamefont
      {Demaille}, \citenamefont {Roditchev}, \citenamefont {Simon},\ and\
      \citenamefont {Cren}}]{Menard2017}%
      \BibitemOpen
      \bibfield  {author} {\bibinfo {author} {\bibfnamefont {G.~C.}\ \bibnamefont
      {M{\'e}nard}}, \bibinfo {author} {\bibfnamefont {S.}~\bibnamefont
      {Guissart}}, \bibinfo {author} {\bibfnamefont {C.}~\bibnamefont {Brun}},
      \bibinfo {author} {\bibfnamefont {R.~T.}\ \bibnamefont {Leriche}}, \bibinfo
      {author} {\bibfnamefont {M.}~\bibnamefont {Trif}}, \bibinfo {author}
      {\bibfnamefont {F.}~\bibnamefont {Debontridder}}, \bibinfo {author}
      {\bibfnamefont {D.}~\bibnamefont {Demaille}}, \bibinfo {author}
      {\bibfnamefont {D.}~\bibnamefont {Roditchev}}, \bibinfo {author}
      {\bibfnamefont {P.}~\bibnamefont {Simon}},\ and\ \bibinfo {author}
      {\bibfnamefont {T.}~\bibnamefont {Cren}},\ }\bibfield  {title} {\bibinfo
      {title} {{Two-dimensional topological superconductivity in Pb/Co/Si(111)}},\
      }\href {https://doi.org/10.1038/s41467-017-02192-x} {\bibfield  {journal}
      {\bibinfo  {journal} {Nat. Commun.}\ }\textbf {\bibinfo {volume} {8}},\
      \bibinfo {pages} {2040} (\bibinfo {year} {2017})}\BibitemShut {NoStop}%
    \bibitem [{\citenamefont {Kezilebieke}\ \emph {et~al.}(2020)\citenamefont
      {Kezilebieke}, \citenamefont {Huda}, \citenamefont {Vaňo}, \citenamefont
      {Aapro}, \citenamefont {Ganguli}, \citenamefont {Silveira}, \citenamefont
      {Głodzik}, \citenamefont {Foster}, \citenamefont {Ojanen},\ and\
      \citenamefont {Liljeroth}}]{Kezilebieke2020}%
      \BibitemOpen
      \bibfield  {author} {\bibinfo {author} {\bibfnamefont {S.}~\bibnamefont
      {Kezilebieke}}, \bibinfo {author} {\bibfnamefont {M.~N.}\ \bibnamefont
      {Huda}}, \bibinfo {author} {\bibfnamefont {V.}~\bibnamefont {Vaňo}},
      \bibinfo {author} {\bibfnamefont {M.}~\bibnamefont {Aapro}}, \bibinfo
      {author} {\bibfnamefont {S.~C.}\ \bibnamefont {Ganguli}}, \bibinfo {author}
      {\bibfnamefont {O.~J.}\ \bibnamefont {Silveira}}, \bibinfo {author}
      {\bibfnamefont {S.}~\bibnamefont {Głodzik}}, \bibinfo {author}
      {\bibfnamefont {A.~S.}\ \bibnamefont {Foster}}, \bibinfo {author}
      {\bibfnamefont {T.}~\bibnamefont {Ojanen}},\ and\ \bibinfo {author}
      {\bibfnamefont {P.}~\bibnamefont {Liljeroth}},\ }\bibfield  {title} {\bibinfo
      {title} {{Topological superconductivity in a van der Waals
      heterostructure}},\ }\href {https://doi.org/10.1038/s41586-020-2989-y}
      {\bibfield  {journal} {\bibinfo  {journal} {Nature}\ }\textbf {\bibinfo
      {volume} {588}},\ \bibinfo {pages} {424} (\bibinfo {year}
      {2020})}\BibitemShut {NoStop}%
    \bibitem [{\citenamefont {Bazarnik}\ \emph {et~al.}(2023)\citenamefont
      {Bazarnik}, \citenamefont {Lo~Conte}, \citenamefont {Mascot}, \citenamefont
      {von Bergmann}, \citenamefont {Morr},\ and\ \citenamefont
      {Wiesendanger}}]{Bazarnik2023}%
      \BibitemOpen
      \bibfield  {author} {\bibinfo {author} {\bibfnamefont {M.}~\bibnamefont
      {Bazarnik}}, \bibinfo {author} {\bibfnamefont {R.}~\bibnamefont {Lo~Conte}},
      \bibinfo {author} {\bibfnamefont {E.}~\bibnamefont {Mascot}}, \bibinfo
      {author} {\bibfnamefont {K.}~\bibnamefont {von Bergmann}}, \bibinfo {author}
      {\bibfnamefont {D.~K.}\ \bibnamefont {Morr}},\ and\ \bibinfo {author}
      {\bibfnamefont {R.}~\bibnamefont {Wiesendanger}},\ }\bibfield  {title}
      {\bibinfo {title} {Antiferromagnetism-driven two-dimensional topological
      nodal-point superconductivity},\ }\href
      {https://doi.org/10.1038/s41467-023-36201-z} {\bibfield  {journal} {\bibinfo
      {journal} {Nat. Commun.}\ }\textbf {\bibinfo {volume} {14}},\ \bibinfo
      {pages} {614} (\bibinfo {year} {2023})}\BibitemShut {NoStop}%
    \bibitem [{\citenamefont {{Deutsch}}\ and\ \citenamefont
      {{Jozsa}}(1992)}]{Deutsch_1992}%
      \BibitemOpen
      \bibfield  {author} {\bibinfo {author} {\bibfnamefont {D.}~\bibnamefont
      {{Deutsch}}}\ and\ \bibinfo {author} {\bibfnamefont {R.}~\bibnamefont
      {{Jozsa}}},\ }\bibfield  {title} {\bibinfo {title} {{Rapid Solution of
      Problems by Quantum Computation}},\ }\href
      {https://doi.org/10.1098/rspa.1992.0167} {\bibfield  {journal} {\bibinfo
      {journal} {Proc. R. Soc. Lond. A}\ }\textbf {\bibinfo {volume} {439}},\
      \bibinfo {pages} {553} (\bibinfo {year} {1992})}\BibitemShut {NoStop}%
    \bibitem [{\citenamefont {Berthiaume}\ and\ \citenamefont
      {Brassard}(1992)}]{Berthiaume_1992}%
      \BibitemOpen
      \bibfield  {author} {\bibinfo {author} {\bibfnamefont {A.}~\bibnamefont
      {Berthiaume}}\ and\ \bibinfo {author} {\bibfnamefont {G.}~\bibnamefont
      {Brassard}},\ }\bibfield  {title} {\bibinfo {title} {Oracle quantum
      computing},\ }in\ \href {https://doi.org/10.1109/PHYCMP.1992.615538} {\emph
      {\bibinfo {booktitle} {Workshop on Physics and Computation}}}\ (\bibinfo
      {publisher} {IEEE Computer Society},\ \bibinfo {year} {1992})\ pp.\ \bibinfo
      {pages} {195--199}\BibitemShut {NoStop}%
    \bibitem [{\citenamefont {Bernstein}\ and\ \citenamefont
      {Vazirani}(1997)}]{Bernstein_1997}%
      \BibitemOpen
      \bibfield  {author} {\bibinfo {author} {\bibfnamefont {E.}~\bibnamefont
      {Bernstein}}\ and\ \bibinfo {author} {\bibfnamefont {U.}~\bibnamefont
      {Vazirani}},\ }\bibfield  {title} {\bibinfo {title} {Quantum complexity
      theory},\ }\href {https://doi.org/10.1137/S0097539796300921} {\bibfield
      {journal} {\bibinfo  {journal} {SIAM J. Comput.}\ }\textbf {\bibinfo {volume}
      {26}},\ \bibinfo {pages} {1411} (\bibinfo {year} {1997})}\BibitemShut
      {NoStop}%
    \bibitem [{\citenamefont {Grover}(1996)}]{Grover_1996}%
      \BibitemOpen
      \bibfield  {author} {\bibinfo {author} {\bibfnamefont {L.~K.}\ \bibnamefont
      {Grover}},\ }\bibfield  {title} {\bibinfo {title} {A fast quantum mechanical
      algorithm for database search},\ }in\ \href
      {https://doi.org/10.1145/237814.237866} {\emph {\bibinfo {booktitle}
      {Proceedings of the Twenty-Eighth Annual ACM Symposium on Theory of
      Computing}}},\ \bibinfo {series and number} {STOC '96}\ (\bibinfo
      {publisher} {Association for Computing Machinery},\ \bibinfo {address} {New
      York, NY, USA},\ \bibinfo {year} {1996})\ pp.\ \bibinfo {pages}
      {212--219}\BibitemShut {NoStop}%
    \bibitem [{\citenamefont {Shor}(1999)}]{Shor_1999}%
      \BibitemOpen
      \bibfield  {author} {\bibinfo {author} {\bibfnamefont {P.~W.}\ \bibnamefont
      {Shor}},\ }\bibfield  {title} {\bibinfo {title} {Polynomial-time algorithms
      for prime factorization and discrete logarithms on a quantum computer},\
      }\href {https://doi.org/10.1137/S0036144598347011} {\bibfield  {journal}
      {\bibinfo  {journal} {SIAM Rev.}\ }\textbf {\bibinfo {volume} {41}},\
      \bibinfo {pages} {303} (\bibinfo {year} {1999})}\BibitemShut {NoStop}%
    \bibitem [{\citenamefont {Kraus}\ \emph {et~al.}(2013)\citenamefont {Kraus},
      \citenamefont {Zoller},\ and\ \citenamefont {Baranov}}]{Kraus_2013}%
      \BibitemOpen
      \bibfield  {author} {\bibinfo {author} {\bibfnamefont {C.~V.}\ \bibnamefont
      {Kraus}}, \bibinfo {author} {\bibfnamefont {P.}~\bibnamefont {Zoller}},\ and\
      \bibinfo {author} {\bibfnamefont {M.~A.}\ \bibnamefont {Baranov}},\
      }\bibfield  {title} {\bibinfo {title} {Braiding of {{Atomic Majorana
      Fermions}} in {{Wire Networks}} and {{Implementation}} of the {{Deutsch-Jozsa
      Algorithm}}},\ }\href {https://doi.org/10.1103/PhysRevLett.111.203001}
      {\bibfield  {journal} {\bibinfo  {journal} {Phys. Rev. Lett.}\ }\textbf
      {\bibinfo {volume} {111}},\ \bibinfo {pages} {203001} (\bibinfo {year}
      {2013})}\BibitemShut {NoStop}%
    \bibitem [{\citenamefont {Alicea}\ \emph {et~al.}(2011)\citenamefont {Alicea},
      \citenamefont {Oreg}, \citenamefont {Refael}, \citenamefont {von Oppen},\
      and\ \citenamefont {Fisher}}]{Alicea_2011}%
      \BibitemOpen
      \bibfield  {author} {\bibinfo {author} {\bibfnamefont {J.}~\bibnamefont
      {Alicea}}, \bibinfo {author} {\bibfnamefont {Y.}~\bibnamefont {Oreg}},
      \bibinfo {author} {\bibfnamefont {G.}~\bibnamefont {Refael}}, \bibinfo
      {author} {\bibfnamefont {F.}~\bibnamefont {von Oppen}},\ and\ \bibinfo
      {author} {\bibfnamefont {M.~P.~A.}\ \bibnamefont {Fisher}},\ }\bibfield
      {title} {\bibinfo {title} {Non-abelian statistics and topological quantum
      information processing in 1d wire networks},\ }\href
      {https://doi.org/10.1038/nphys1915} {\bibfield  {journal} {\bibinfo
      {journal} {Nat. Phys.}\ }\textbf {\bibinfo {volume} {7}},\ \bibinfo {pages}
      {412} (\bibinfo {year} {2011})}\BibitemShut {NoStop}%
    \bibitem [{\citenamefont {Halperin}\ \emph {et~al.}(2012)\citenamefont
      {Halperin}, \citenamefont {Oreg}, \citenamefont {Stern}, \citenamefont
      {Refael}, \citenamefont {Alicea},\ and\ \citenamefont {{von
      Oppen}}}]{Halperin_2012}%
      \BibitemOpen
      \bibfield  {author} {\bibinfo {author} {\bibfnamefont {B.~I.}\ \bibnamefont
      {Halperin}}, \bibinfo {author} {\bibfnamefont {Y.}~\bibnamefont {Oreg}},
      \bibinfo {author} {\bibfnamefont {A.}~\bibnamefont {Stern}}, \bibinfo
      {author} {\bibfnamefont {G.}~\bibnamefont {Refael}}, \bibinfo {author}
      {\bibfnamefont {J.}~\bibnamefont {Alicea}},\ and\ \bibinfo {author}
      {\bibfnamefont {F.}~\bibnamefont {{von Oppen}}},\ }\bibfield  {title}
      {\bibinfo {title} {Adiabatic manipulations of {{Majorana}} fermions in a
      three-dimensional network of quantum wires},\ }\href
      {https://doi.org/10.1103/PhysRevB.85.144501} {\bibfield  {journal} {\bibinfo
      {journal} {Phys. Rev. B}\ }\textbf {\bibinfo {volume} {85}},\ \bibinfo
      {pages} {144501} (\bibinfo {year} {2012})}\BibitemShut {NoStop}%
    \bibitem [{\citenamefont {Sekania}\ \emph {et~al.}(2017)\citenamefont
      {Sekania}, \citenamefont {Plugge}, \citenamefont {Greiter}, \citenamefont
      {Thomale},\ and\ \citenamefont {Schmitteckert}}]{Sekania_2017}%
      \BibitemOpen
      \bibfield  {author} {\bibinfo {author} {\bibfnamefont {M.}~\bibnamefont
      {Sekania}}, \bibinfo {author} {\bibfnamefont {S.}~\bibnamefont {Plugge}},
      \bibinfo {author} {\bibfnamefont {M.}~\bibnamefont {Greiter}}, \bibinfo
      {author} {\bibfnamefont {R.}~\bibnamefont {Thomale}},\ and\ \bibinfo {author}
      {\bibfnamefont {P.}~\bibnamefont {Schmitteckert}},\ }\bibfield  {title}
      {\bibinfo {title} {Braiding errors in interacting {{Majorana}} quantum
      wires},\ }\href {https://doi.org/10.1103/PhysRevB.96.094307} {\bibfield
      {journal} {\bibinfo  {journal} {Phys. Rev. B}\ }\textbf {\bibinfo {volume}
      {96}},\ \bibinfo {pages} {094307} (\bibinfo {year} {2017})}\BibitemShut
      {NoStop}%
    \bibitem [{\citenamefont {Harper}\ \emph {et~al.}(2019)\citenamefont {Harper},
      \citenamefont {Pushp},\ and\ \citenamefont {Roy}}]{Harper_2019}%
      \BibitemOpen
      \bibfield  {author} {\bibinfo {author} {\bibfnamefont {F.}~\bibnamefont
      {Harper}}, \bibinfo {author} {\bibfnamefont {A.}~\bibnamefont {Pushp}},\ and\
      \bibinfo {author} {\bibfnamefont {R.}~\bibnamefont {Roy}},\ }\bibfield
      {title} {\bibinfo {title} {Majorana braiding in realistic nanowire
      {{Y-junctions}} and tuning forks},\ }\href
      {https://doi.org/10.1103/PhysRevResearch.1.033207} {\bibfield  {journal}
      {\bibinfo  {journal} {Phys. Rev. Res.}\ }\textbf {\bibinfo {volume} {1}},\
      \bibinfo {pages} {033207} (\bibinfo {year} {2019})}\BibitemShut {NoStop}%
    \bibitem [{\citenamefont {Tutschku}\ \emph {et~al.}(2020)\citenamefont
      {Tutschku}, \citenamefont {Reinthaler}, \citenamefont {Lei}, \citenamefont
      {MacDonald},\ and\ \citenamefont {Hankiewicz}}]{Tutschku_2020}%
      \BibitemOpen
      \bibfield  {author} {\bibinfo {author} {\bibfnamefont {C.}~\bibnamefont
      {Tutschku}}, \bibinfo {author} {\bibfnamefont {R.~W.}\ \bibnamefont
      {Reinthaler}}, \bibinfo {author} {\bibfnamefont {C.}~\bibnamefont {Lei}},
      \bibinfo {author} {\bibfnamefont {A.~H.}\ \bibnamefont {MacDonald}},\ and\
      \bibinfo {author} {\bibfnamefont {E.~M.}\ \bibnamefont {Hankiewicz}},\
      }\bibfield  {title} {\bibinfo {title} {Majorana-based quantum computing in
      nanowire devices},\ }\href {https://doi.org/10.1103/PhysRevB.102.125407}
      {\bibfield  {journal} {\bibinfo  {journal} {Phys. Rev. B}\ }\textbf {\bibinfo
      {volume} {102}},\ \bibinfo {pages} {125407} (\bibinfo {year}
      {2020})}\BibitemShut {NoStop}%
    \bibitem [{\citenamefont {Mascot}\ \emph {et~al.}(2023)\citenamefont {Mascot},
      \citenamefont {Hodge}, \citenamefont {Crawford}, \citenamefont {Bedow},
      \citenamefont {Morr},\ and\ \citenamefont {Rachel}}]{Mascot_2023}%
      \BibitemOpen
      \bibfield  {author} {\bibinfo {author} {\bibfnamefont {E.}~\bibnamefont
      {Mascot}}, \bibinfo {author} {\bibfnamefont {T.}~\bibnamefont {Hodge}},
      \bibinfo {author} {\bibfnamefont {D.}~\bibnamefont {Crawford}}, \bibinfo
      {author} {\bibfnamefont {J.}~\bibnamefont {Bedow}}, \bibinfo {author}
      {\bibfnamefont {D.~K.}\ \bibnamefont {Morr}},\ and\ \bibinfo {author}
      {\bibfnamefont {S.}~\bibnamefont {Rachel}},\ }\bibfield  {title} {\bibinfo
      {title} {Many-body majorana braiding without an exponential hilbert space},\
      }\href {https://doi.org/10.1103/PhysRevLett.131.176601} {\bibfield  {journal}
      {\bibinfo  {journal} {Phys. Rev. Lett.}\ }\textbf {\bibinfo {volume} {131}},\
      \bibinfo {pages} {176601} (\bibinfo {year} {2023})}\BibitemShut {NoStop}%
    \bibitem [{\citenamefont {Hodge}\ \emph {et~al.}(2025)\citenamefont {Hodge},
      \citenamefont {Mascot}, \citenamefont {Crawford},\ and\ \citenamefont
      {Rachel}}]{Hodge_2025}%
      \BibitemOpen
      \bibfield  {author} {\bibinfo {author} {\bibfnamefont {T.}~\bibnamefont
      {Hodge}}, \bibinfo {author} {\bibfnamefont {E.}~\bibnamefont {Mascot}},
      \bibinfo {author} {\bibfnamefont {D.}~\bibnamefont {Crawford}},\ and\
      \bibinfo {author} {\bibfnamefont {S.}~\bibnamefont {Rachel}},\ }\bibfield
      {title} {\bibinfo {title} {Characterizing dynamic hybridization of majorana
      zero modes for universal quantum computing},\ }\href
      {https://doi.org/10.1103/PhysRevLett.134.096601} {\bibfield  {journal}
      {\bibinfo  {journal} {Phys. Rev. Lett.}\ }\textbf {\bibinfo {volume} {134}},\
      \bibinfo {pages} {096601} (\bibinfo {year} {2025})}\BibitemShut {NoStop}%
    \bibitem [{\citenamefont {Amorim}\ \emph {et~al.}(2015)\citenamefont {Amorim},
      \citenamefont {Ebihara}, \citenamefont {Yamakage}, \citenamefont {Tanaka},\
      and\ \citenamefont {Sato}}]{Amorim_2015}%
      \BibitemOpen
      \bibfield  {author} {\bibinfo {author} {\bibfnamefont {C.~S.}\ \bibnamefont
      {Amorim}}, \bibinfo {author} {\bibfnamefont {K.}~\bibnamefont {Ebihara}},
      \bibinfo {author} {\bibfnamefont {A.}~\bibnamefont {Yamakage}}, \bibinfo
      {author} {\bibfnamefont {Y.}~\bibnamefont {Tanaka}},\ and\ \bibinfo {author}
      {\bibfnamefont {M.}~\bibnamefont {Sato}},\ }\bibfield  {title} {\bibinfo
      {title} {Majorana braiding dynamics in nanowires},\ }\href
      {https://doi.org/10.1103/PhysRevB.91.174305} {\bibfield  {journal} {\bibinfo
      {journal} {Phys. Rev. B}\ }\textbf {\bibinfo {volume} {91}},\ \bibinfo
      {pages} {174305} (\bibinfo {year} {2015})}\BibitemShut {NoStop}%
    \bibitem [{\citenamefont {Aasen}\ \emph {et~al.}(2016)\citenamefont {Aasen},
      \citenamefont {Hell}, \citenamefont {Mishmash}, \citenamefont {Higginbotham},
      \citenamefont {Danon}, \citenamefont {Leijnse}, \citenamefont {Jespersen},
      \citenamefont {Folk}, \citenamefont {Marcus}, \citenamefont {Flensberg},\
      and\ \citenamefont {Alicea}}]{Aasen_2016}%
      \BibitemOpen
      \bibfield  {author} {\bibinfo {author} {\bibfnamefont {D.}~\bibnamefont
      {Aasen}}, \bibinfo {author} {\bibfnamefont {M.}~\bibnamefont {Hell}},
      \bibinfo {author} {\bibfnamefont {R.~V.}\ \bibnamefont {Mishmash}}, \bibinfo
      {author} {\bibfnamefont {A.}~\bibnamefont {Higginbotham}}, \bibinfo {author}
      {\bibfnamefont {J.}~\bibnamefont {Danon}}, \bibinfo {author} {\bibfnamefont
      {M.}~\bibnamefont {Leijnse}}, \bibinfo {author} {\bibfnamefont {T.~S.}\
      \bibnamefont {Jespersen}}, \bibinfo {author} {\bibfnamefont {J.~A.}\
      \bibnamefont {Folk}}, \bibinfo {author} {\bibfnamefont {C.~M.}\ \bibnamefont
      {Marcus}}, \bibinfo {author} {\bibfnamefont {K.}~\bibnamefont {Flensberg}},\
      and\ \bibinfo {author} {\bibfnamefont {J.}~\bibnamefont {Alicea}},\
      }\bibfield  {title} {\bibinfo {title} {Milestones toward majorana-based
      quantum computing},\ }\href {https://doi.org/10.1103/PhysRevX.6.031016}
      {\bibfield  {journal} {\bibinfo  {journal} {Phys. Rev. X}\ }\textbf {\bibinfo
      {volume} {6}},\ \bibinfo {pages} {031016} (\bibinfo {year}
      {2016})}\BibitemShut {NoStop}%
    \bibitem [{\citenamefont {Karzig}\ \emph {et~al.}(2017)\citenamefont {Karzig},
      \citenamefont {Knapp}, \citenamefont {Lutchyn}, \citenamefont {Bonderson},
      \citenamefont {Hastings}, \citenamefont {Nayak}, \citenamefont {Alicea},
      \citenamefont {Flensberg}, \citenamefont {Plugge}, \citenamefont {Oreg},
      \citenamefont {Marcus},\ and\ \citenamefont {Freedman}}]{Karzig_2017}%
      \BibitemOpen
      \bibfield  {author} {\bibinfo {author} {\bibfnamefont {T.}~\bibnamefont
      {Karzig}}, \bibinfo {author} {\bibfnamefont {C.}~\bibnamefont {Knapp}},
      \bibinfo {author} {\bibfnamefont {R.~M.}\ \bibnamefont {Lutchyn}}, \bibinfo
      {author} {\bibfnamefont {P.}~\bibnamefont {Bonderson}}, \bibinfo {author}
      {\bibfnamefont {M.~B.}\ \bibnamefont {Hastings}}, \bibinfo {author}
      {\bibfnamefont {C.}~\bibnamefont {Nayak}}, \bibinfo {author} {\bibfnamefont
      {J.}~\bibnamefont {Alicea}}, \bibinfo {author} {\bibfnamefont
      {K.}~\bibnamefont {Flensberg}}, \bibinfo {author} {\bibfnamefont
      {S.}~\bibnamefont {Plugge}}, \bibinfo {author} {\bibfnamefont
      {Y.}~\bibnamefont {Oreg}}, \bibinfo {author} {\bibfnamefont {C.~M.}\
      \bibnamefont {Marcus}},\ and\ \bibinfo {author} {\bibfnamefont {M.~H.}\
      \bibnamefont {Freedman}},\ }\bibfield  {title} {\bibinfo {title} {Scalable
      designs for quasiparticle-poisoning-protected topological quantum computation
      with {{Majorana}} zero modes},\ }\href
      {https://doi.org/10.1103/PhysRevB.95.235305} {\bibfield  {journal} {\bibinfo
      {journal} {Phys. Rev. B}\ }\textbf {\bibinfo {volume} {95}},\ \bibinfo
      {pages} {235305} (\bibinfo {year} {2017})}\BibitemShut {NoStop}%
    \bibitem [{\citenamefont {Zhou}\ \emph {et~al.}(2022)\citenamefont {Zhou},
      \citenamefont {Dartiailh}, \citenamefont {Sardashti}, \citenamefont {Han},
      \citenamefont {{Matos-Abiague}}, \citenamefont {Shabani},\ and\ \citenamefont
      {{\v Z}uti{\'c}}}]{Zhou_2022}%
      \BibitemOpen
      \bibfield  {author} {\bibinfo {author} {\bibfnamefont {T.}~\bibnamefont
      {Zhou}}, \bibinfo {author} {\bibfnamefont {M.~C.}\ \bibnamefont {Dartiailh}},
      \bibinfo {author} {\bibfnamefont {K.}~\bibnamefont {Sardashti}}, \bibinfo
      {author} {\bibfnamefont {J.~E.}\ \bibnamefont {Han}}, \bibinfo {author}
      {\bibfnamefont {A.}~\bibnamefont {{Matos-Abiague}}}, \bibinfo {author}
      {\bibfnamefont {J.}~\bibnamefont {Shabani}},\ and\ \bibinfo {author}
      {\bibfnamefont {I.}~\bibnamefont {{\v Z}uti{\'c}}},\ }\bibfield  {title}
      {\bibinfo {title} {Fusion of {{Majorana}} bound states with mini-gate control
      in two-dimensional systems},\ }\href
      {https://doi.org/10.1038/s41467-022-29463-6} {\bibfield  {journal} {\bibinfo
      {journal} {Nat. Commun.}\ }\textbf {\bibinfo {volume} {13}},\ \bibinfo
      {pages} {1738} (\bibinfo {year} {2022})}\BibitemShut {NoStop}%
    \bibitem [{\citenamefont {Sanno}\ \emph {et~al.}(2021)\citenamefont {Sanno},
      \citenamefont {Miyazaki}, \citenamefont {Mizushima},\ and\ \citenamefont
      {Fujimoto}}]{Sanno_2021}%
      \BibitemOpen
      \bibfield  {author} {\bibinfo {author} {\bibfnamefont {T.}~\bibnamefont
      {Sanno}}, \bibinfo {author} {\bibfnamefont {S.}~\bibnamefont {Miyazaki}},
      \bibinfo {author} {\bibfnamefont {T.}~\bibnamefont {Mizushima}},\ and\
      \bibinfo {author} {\bibfnamefont {S.}~\bibnamefont {Fujimoto}},\ }\bibfield
      {title} {\bibinfo {title} {{\emph{Ab Initio}} simulation of non-{{Abelian}}
      braiding statistics in topological superconductors},\ }\href
      {https://doi.org/10.1103/PhysRevB.103.054504} {\bibfield  {journal} {\bibinfo
       {journal} {Phys. Rev. B}\ }\textbf {\bibinfo {volume} {103}},\ \bibinfo
      {pages} {054504} (\bibinfo {year} {2021})}\BibitemShut {NoStop}%
    \bibitem [{\citenamefont {Hyart}\ \emph {et~al.}(2013)\citenamefont {Hyart},
      \citenamefont {van Heck}, \citenamefont {Fulga}, \citenamefont {Burrello},
      \citenamefont {Akhmerov},\ and\ \citenamefont {Beenakker}}]{Hyart_2013}%
      \BibitemOpen
      \bibfield  {author} {\bibinfo {author} {\bibfnamefont {T.}~\bibnamefont
      {Hyart}}, \bibinfo {author} {\bibfnamefont {B.}~\bibnamefont {van Heck}},
      \bibinfo {author} {\bibfnamefont {I.~C.}\ \bibnamefont {Fulga}}, \bibinfo
      {author} {\bibfnamefont {M.}~\bibnamefont {Burrello}}, \bibinfo {author}
      {\bibfnamefont {A.~R.}\ \bibnamefont {Akhmerov}},\ and\ \bibinfo {author}
      {\bibfnamefont {C.~W.~J.}\ \bibnamefont {Beenakker}},\ }\bibfield  {title}
      {\bibinfo {title} {Flux-controlled quantum computation with majorana
      fermions},\ }\href {https://doi.org/10.1103/PhysRevB.88.035121} {\bibfield
      {journal} {\bibinfo  {journal} {Phys. Rev. B}\ }\textbf {\bibinfo {volume}
      {88}},\ \bibinfo {pages} {035121} (\bibinfo {year} {2013})}\BibitemShut
      {NoStop}%
    \bibitem [{\citenamefont {Li}\ \emph {et~al.}(2016)\citenamefont {Li},
      \citenamefont {Neupert}, \citenamefont {Bernevig},\ and\ \citenamefont
      {Yazdani}}]{Li_2016}%
      \BibitemOpen
      \bibfield  {author} {\bibinfo {author} {\bibfnamefont {J.}~\bibnamefont
      {Li}}, \bibinfo {author} {\bibfnamefont {T.}~\bibnamefont {Neupert}},
      \bibinfo {author} {\bibfnamefont {B.~A.}\ \bibnamefont {Bernevig}},\ and\
      \bibinfo {author} {\bibfnamefont {A.}~\bibnamefont {Yazdani}},\ }\bibfield
      {title} {\bibinfo {title} {Manipulating {{Majorana}} zero modes on atomic
      rings with an external magnetic field},\ }\href
      {https://doi.org/10.1038/ncomms10395} {\bibfield  {journal} {\bibinfo
      {journal} {Nat. Commun.}\ }\textbf {\bibinfo {volume} {7}},\ \bibinfo {pages}
      {10395} (\bibinfo {year} {2016})}\BibitemShut {NoStop}%
    \bibitem [{\citenamefont {Bedow}\ \emph {et~al.}(2024)\citenamefont {Bedow},
      \citenamefont {Mascot}, \citenamefont {Hodge}, \citenamefont {Rachel},\ and\
      \citenamefont {Morr}}]{Bedow2024}%
      \BibitemOpen
      \bibfield  {author} {\bibinfo {author} {\bibfnamefont {J.}~\bibnamefont
      {Bedow}}, \bibinfo {author} {\bibfnamefont {E.}~\bibnamefont {Mascot}},
      \bibinfo {author} {\bibfnamefont {T.}~\bibnamefont {Hodge}}, \bibinfo
      {author} {\bibfnamefont {S.}~\bibnamefont {Rachel}},\ and\ \bibinfo {author}
      {\bibfnamefont {D.~K.}\ \bibnamefont {Morr}},\ }\bibfield  {title} {\bibinfo
      {title} {Simulating topological quantum gates in two-dimensional
      magnet-superconductor hybrid structures},\ }\href
      {https://doi.org/10.1038/s41535-024-00703-w} {\bibfield  {journal} {\bibinfo
      {journal} {npj Quantum Mater.}\ }\textbf {\bibinfo {volume} {9}},\ \bibinfo
      {pages} {99} (\bibinfo {year} {2024})}\BibitemShut {NoStop}%
    \bibitem [{\citenamefont {Du}\ \emph {et~al.}(2001)\citenamefont {Du},
      \citenamefont {Shi}, \citenamefont {Zhou}, \citenamefont {Fan}, \citenamefont
      {Ye}, \citenamefont {Han},\ and\ \citenamefont {Wu}}]{Du_2001}%
      \BibitemOpen
      \bibfield  {author} {\bibinfo {author} {\bibfnamefont {J.}~\bibnamefont
      {Du}}, \bibinfo {author} {\bibfnamefont {M.}~\bibnamefont {Shi}}, \bibinfo
      {author} {\bibfnamefont {X.}~\bibnamefont {Zhou}}, \bibinfo {author}
      {\bibfnamefont {Y.}~\bibnamefont {Fan}}, \bibinfo {author} {\bibfnamefont
      {B.}~\bibnamefont {Ye}}, \bibinfo {author} {\bibfnamefont {R.}~\bibnamefont
      {Han}},\ and\ \bibinfo {author} {\bibfnamefont {J.}~\bibnamefont {Wu}},\
      }\bibfield  {title} {\bibinfo {title} {Implementation of a quantum algorithm
      to solve the {{Bernstein-Vazirani}} parity problem without entanglement on an
      ensemble quantum computer},\ }\href
      {https://doi.org/10.1103/PhysRevA.64.042306} {\bibfield  {journal} {\bibinfo
      {journal} {Phys. Rev. A}\ }\textbf {\bibinfo {volume} {64}},\ \bibinfo
      {pages} {042306} (\bibinfo {year} {2001})}\BibitemShut {NoStop}%
    \bibitem [{\citenamefont {Brainis}\ \emph {et~al.}(2003)\citenamefont
      {Brainis}, \citenamefont {Lamoureux}, \citenamefont {Cerf}, \citenamefont
      {Emplit}, \citenamefont {Haelterman},\ and\ \citenamefont
      {Massar}}]{Brainis_2003}%
      \BibitemOpen
      \bibfield  {author} {\bibinfo {author} {\bibfnamefont {E.}~\bibnamefont
      {Brainis}}, \bibinfo {author} {\bibfnamefont {L.-P.}\ \bibnamefont
      {Lamoureux}}, \bibinfo {author} {\bibfnamefont {N.~J.}\ \bibnamefont {Cerf}},
      \bibinfo {author} {\bibfnamefont {{\relax Ph}.}~\bibnamefont {Emplit}},
      \bibinfo {author} {\bibfnamefont {M.}~\bibnamefont {Haelterman}},\ and\
      \bibinfo {author} {\bibfnamefont {S.}~\bibnamefont {Massar}},\ }\bibfield
      {title} {\bibinfo {title} {Fiber-{{Optics Implementation}} of the
      {{Deutsch-Jozsa}} and {{Bernstein-Vazirani Quantum Algorithms}} with {{Three
      Qubits}}},\ }\href {https://doi.org/10.1103/PhysRevLett.90.157902} {\bibfield
       {journal} {\bibinfo  {journal} {Phys. Rev. Lett.}\ }\textbf {\bibinfo
      {volume} {90}},\ \bibinfo {pages} {157902} (\bibinfo {year}
      {2003})}\BibitemShut {NoStop}%
    \bibitem [{\citenamefont {Londero}\ \emph {et~al.}(2004)\citenamefont
      {Londero}, \citenamefont {Dorrer}, \citenamefont {Anderson}, \citenamefont
      {Wallentowitz}, \citenamefont {Banaszek},\ and\ \citenamefont
      {Walmsley}}]{Londero_2004}%
      \BibitemOpen
      \bibfield  {author} {\bibinfo {author} {\bibfnamefont {P.}~\bibnamefont
      {Londero}}, \bibinfo {author} {\bibfnamefont {C.}~\bibnamefont {Dorrer}},
      \bibinfo {author} {\bibfnamefont {M.}~\bibnamefont {Anderson}}, \bibinfo
      {author} {\bibfnamefont {S.}~\bibnamefont {Wallentowitz}}, \bibinfo {author}
      {\bibfnamefont {K.}~\bibnamefont {Banaszek}},\ and\ \bibinfo {author}
      {\bibfnamefont {I.~A.}\ \bibnamefont {Walmsley}},\ }\bibfield  {title}
      {\bibinfo {title} {Efficient optical implementation of the
      {{Bernstein-Vazirani}} algorithm},\ }\href
      {https://doi.org/10.1103/PhysRevA.69.010302} {\bibfield  {journal} {\bibinfo
      {journal} {Phys. Rev. A}\ }\textbf {\bibinfo {volume} {69}},\ \bibinfo
      {pages} {010302} (\bibinfo {year} {2004})}\BibitemShut {NoStop}%
    \bibitem [{\citenamefont {Peng}\ \emph {et~al.}(2004)\citenamefont {Peng},
      \citenamefont {Zhu}, \citenamefont {Fang}, \citenamefont {Feng},
      \citenamefont {Liu},\ and\ \citenamefont {Gao}}]{Peng_2004}%
      \BibitemOpen
      \bibfield  {author} {\bibinfo {author} {\bibfnamefont {X.}~\bibnamefont
      {Peng}}, \bibinfo {author} {\bibfnamefont {X.}~\bibnamefont {Zhu}}, \bibinfo
      {author} {\bibfnamefont {X.}~\bibnamefont {Fang}}, \bibinfo {author}
      {\bibfnamefont {M.}~\bibnamefont {Feng}}, \bibinfo {author} {\bibfnamefont
      {M.}~\bibnamefont {Liu}},\ and\ \bibinfo {author} {\bibfnamefont
      {K.}~\bibnamefont {Gao}},\ }\bibfield  {title} {\bibinfo {title}
      {``{{Spectral}} implementation'' for creating a labeled pseudo-pure state and
      the {{Bernstein}}--{{Vazirani}} algorithm in a four-qubit nuclear magnetic
      resonance quantum processor},\ }\href {https://doi.org/10.1063/1.1642579}
      {\bibfield  {journal} {\bibinfo  {journal} {J. Chem. Phys.}\ }\textbf
      {\bibinfo {volume} {120}},\ \bibinfo {pages} {3579} (\bibinfo {year}
      {2004})}\BibitemShut {NoStop}%
    \bibitem [{\citenamefont {Debnath}\ \emph {et~al.}(2016)\citenamefont
      {Debnath}, \citenamefont {Linke}, \citenamefont {Figgatt}, \citenamefont
      {Landsman}, \citenamefont {Wright},\ and\ \citenamefont
      {Monroe}}]{Debnath_2016}%
      \BibitemOpen
      \bibfield  {author} {\bibinfo {author} {\bibfnamefont {S.}~\bibnamefont
      {Debnath}}, \bibinfo {author} {\bibfnamefont {N.~M.}\ \bibnamefont {Linke}},
      \bibinfo {author} {\bibfnamefont {C.}~\bibnamefont {Figgatt}}, \bibinfo
      {author} {\bibfnamefont {K.~A.}\ \bibnamefont {Landsman}}, \bibinfo {author}
      {\bibfnamefont {K.}~\bibnamefont {Wright}},\ and\ \bibinfo {author}
      {\bibfnamefont {C.}~\bibnamefont {Monroe}},\ }\bibfield  {title} {\bibinfo
      {title} {Demonstration of a small programmable quantum computer with atomic
      qubits},\ }\href {https://doi.org/10.1038/nature18648} {\bibfield  {journal}
      {\bibinfo  {journal} {Nature}\ }\textbf {\bibinfo {volume} {536}},\ \bibinfo
      {pages} {63} (\bibinfo {year} {2016})}\BibitemShut {NoStop}%
    \bibitem [{\citenamefont {Wright}\ \emph {et~al.}(2019)\citenamefont {Wright},
      \citenamefont {Beck}, \citenamefont {Debnath}, \citenamefont {Amini},
      \citenamefont {Nam}, \citenamefont {Grzesiak}, \citenamefont {Chen},
      \citenamefont {Pisenti}, \citenamefont {Chmielewski}, \citenamefont
      {Collins}, \citenamefont {Hudek}, \citenamefont {Mizrahi}, \citenamefont
      {{Wong-Campos}}, \citenamefont {Allen}, \citenamefont {Apisdorf},
      \citenamefont {Solomon}, \citenamefont {Williams}, \citenamefont {Ducore},
      \citenamefont {Blinov}, \citenamefont {Kreikemeier}, \citenamefont {Chaplin},
      \citenamefont {Keesan}, \citenamefont {Monroe},\ and\ \citenamefont
      {Kim}}]{Wright_2019}%
      \BibitemOpen
      \bibfield  {author} {\bibinfo {author} {\bibfnamefont {K.}~\bibnamefont
      {Wright}}, \bibinfo {author} {\bibfnamefont {K.~M.}\ \bibnamefont {Beck}},
      \bibinfo {author} {\bibfnamefont {S.}~\bibnamefont {Debnath}}, \bibinfo
      {author} {\bibfnamefont {J.~M.}\ \bibnamefont {Amini}}, \bibinfo {author}
      {\bibfnamefont {Y.}~\bibnamefont {Nam}}, \bibinfo {author} {\bibfnamefont
      {N.}~\bibnamefont {Grzesiak}}, \bibinfo {author} {\bibfnamefont {J.-S.}\
      \bibnamefont {Chen}}, \bibinfo {author} {\bibfnamefont {N.~C.}\ \bibnamefont
      {Pisenti}}, \bibinfo {author} {\bibfnamefont {M.}~\bibnamefont
      {Chmielewski}}, \bibinfo {author} {\bibfnamefont {C.}~\bibnamefont
      {Collins}}, \bibinfo {author} {\bibfnamefont {K.~M.}\ \bibnamefont {Hudek}},
      \bibinfo {author} {\bibfnamefont {J.}~\bibnamefont {Mizrahi}}, \bibinfo
      {author} {\bibfnamefont {J.~D.}\ \bibnamefont {{Wong-Campos}}}, \bibinfo
      {author} {\bibfnamefont {S.}~\bibnamefont {Allen}}, \bibinfo {author}
      {\bibfnamefont {J.}~\bibnamefont {Apisdorf}}, \bibinfo {author}
      {\bibfnamefont {P.}~\bibnamefont {Solomon}}, \bibinfo {author} {\bibfnamefont
      {M.}~\bibnamefont {Williams}}, \bibinfo {author} {\bibfnamefont {A.~M.}\
      \bibnamefont {Ducore}}, \bibinfo {author} {\bibfnamefont {A.}~\bibnamefont
      {Blinov}}, \bibinfo {author} {\bibfnamefont {S.~M.}\ \bibnamefont
      {Kreikemeier}}, \bibinfo {author} {\bibfnamefont {V.}~\bibnamefont
      {Chaplin}}, \bibinfo {author} {\bibfnamefont {M.}~\bibnamefont {Keesan}},
      \bibinfo {author} {\bibfnamefont {C.}~\bibnamefont {Monroe}},\ and\ \bibinfo
      {author} {\bibfnamefont {J.}~\bibnamefont {Kim}},\ }\bibfield  {title}
      {\bibinfo {title} {Benchmarking an 11-qubit quantum computer},\ }\href
      {https://doi.org/10.1038/s41467-019-13534-2} {\bibfield  {journal} {\bibinfo
      {journal} {Nat. Commun.}\ }\textbf {\bibinfo {volume} {10}},\ \bibinfo
      {pages} {5464} (\bibinfo {year} {2019})}\BibitemShut {NoStop}%
    \bibitem [{\citenamefont {Yang}\ \emph {et~al.}(2018)\citenamefont {Yang},
      \citenamefont {Willke}, \citenamefont {Bae}, \citenamefont {Ferr{\'o}n},
      \citenamefont {Lado}, \citenamefont {Ardavan}, \citenamefont
      {{Fern{\'a}ndez-Rossier}}, \citenamefont {Heinrich},\ and\ \citenamefont
      {Lutz}}]{Yang_2018}%
      \BibitemOpen
      \bibfield  {author} {\bibinfo {author} {\bibfnamefont {K.}~\bibnamefont
      {Yang}}, \bibinfo {author} {\bibfnamefont {P.}~\bibnamefont {Willke}},
      \bibinfo {author} {\bibfnamefont {Y.}~\bibnamefont {Bae}}, \bibinfo {author}
      {\bibfnamefont {A.}~\bibnamefont {Ferr{\'o}n}}, \bibinfo {author}
      {\bibfnamefont {J.~L.}\ \bibnamefont {Lado}}, \bibinfo {author}
      {\bibfnamefont {A.}~\bibnamefont {Ardavan}}, \bibinfo {author} {\bibfnamefont
      {J.}~\bibnamefont {{Fern{\'a}ndez-Rossier}}}, \bibinfo {author}
      {\bibfnamefont {A.~J.}\ \bibnamefont {Heinrich}},\ and\ \bibinfo {author}
      {\bibfnamefont {C.~P.}\ \bibnamefont {Lutz}},\ }\bibfield  {title} {\bibinfo
      {title} {Electrically controlled nuclear polarization of individual atoms},\
      }\href {https://doi.org/10.1038/s41565-018-0296-7} {\bibfield  {journal}
      {\bibinfo  {journal} {Nat. Nanotechnol.}\ }\textbf {\bibinfo {volume} {13}},\
      \bibinfo {pages} {1120} (\bibinfo {year} {2018})}\BibitemShut {NoStop}%
    \bibitem [{\citenamefont {Yang}\ \emph {et~al.}(2019)\citenamefont {Yang},
      \citenamefont {Paul}, \citenamefont {Phark}, \citenamefont {Willke},
      \citenamefont {Bae}, \citenamefont {Choi}, \citenamefont {Esat},
      \citenamefont {Ardavan}, \citenamefont {Heinrich},\ and\ \citenamefont
      {Lutz}}]{Yang_2019}%
      \BibitemOpen
      \bibfield  {author} {\bibinfo {author} {\bibfnamefont {K.}~\bibnamefont
      {Yang}}, \bibinfo {author} {\bibfnamefont {W.}~\bibnamefont {Paul}}, \bibinfo
      {author} {\bibfnamefont {S.}~\bibnamefont {Phark}}, \bibinfo {author}
      {\bibfnamefont {P.}~\bibnamefont {Willke}}, \bibinfo {author} {\bibfnamefont
      {Y.}~\bibnamefont {Bae}}, \bibinfo {author} {\bibfnamefont {T.}~\bibnamefont
      {Choi}}, \bibinfo {author} {\bibfnamefont {T.}~\bibnamefont {Esat}}, \bibinfo
      {author} {\bibfnamefont {A.}~\bibnamefont {Ardavan}}, \bibinfo {author}
      {\bibfnamefont {A.~J.}\ \bibnamefont {Heinrich}},\ and\ \bibinfo {author}
      {\bibfnamefont {C.~P.}\ \bibnamefont {Lutz}},\ }\bibfield  {title} {\bibinfo
      {title} {Coherent spin manipulation of individual atoms on a surface},\
      }\href {https://doi.org/10.1126/science.aay6779} {\bibfield  {journal}
      {\bibinfo  {journal} {Science}\ }\textbf {\bibinfo {volume} {366}},\ \bibinfo
      {pages} {509} (\bibinfo {year} {2019})}\BibitemShut {NoStop}%
    \bibitem [{\citenamefont {Phark}\ \emph {et~al.}(2023)\citenamefont {Phark},
      \citenamefont {Bui}, \citenamefont {Ferrón}, \citenamefont
      {Fernández-Rossier}, \citenamefont {Reina-Gálvez}, \citenamefont {Wolf},
      \citenamefont {Wang}, \citenamefont {Yang}, \citenamefont {Heinrich},\ and\
      \citenamefont {Lutz}}]{Phark_2023}%
      \BibitemOpen
      \bibfield  {author} {\bibinfo {author} {\bibfnamefont {S.-h.}\ \bibnamefont
      {Phark}}, \bibinfo {author} {\bibfnamefont {H.~T.}\ \bibnamefont {Bui}},
      \bibinfo {author} {\bibfnamefont {A.}~\bibnamefont {Ferrón}}, \bibinfo
      {author} {\bibfnamefont {J.}~\bibnamefont {Fernández-Rossier}}, \bibinfo
      {author} {\bibfnamefont {J.}~\bibnamefont {Reina-Gálvez}}, \bibinfo {author}
      {\bibfnamefont {C.}~\bibnamefont {Wolf}}, \bibinfo {author} {\bibfnamefont
      {Y.}~\bibnamefont {Wang}}, \bibinfo {author} {\bibfnamefont {K.}~\bibnamefont
      {Yang}}, \bibinfo {author} {\bibfnamefont {A.~J.}\ \bibnamefont {Heinrich}},\
      and\ \bibinfo {author} {\bibfnamefont {C.~P.}\ \bibnamefont {Lutz}},\
      }\bibfield  {title} {\bibinfo {title} {Electric-field-driven spin resonance
      by on-surface exchange coupling to a single-atom magnet},\ }\href
      {https://doi.org/https://doi.org/10.1002/advs.202302033} {\bibfield
      {journal} {\bibinfo  {journal} {Adv. Sci.}\ }\textbf {\bibinfo {volume}
      {10}},\ \bibinfo {pages} {2302033} (\bibinfo {year} {2023})}\BibitemShut
      {NoStop}%
    \bibitem [{\citenamefont {Wang}\ \emph {et~al.}(2023)\citenamefont {Wang},
      \citenamefont {Haze}, \citenamefont {Bui}, \citenamefont {Soe}, \citenamefont
      {Aubin}, \citenamefont {Ardavan}, \citenamefont {Heinrich},\ and\
      \citenamefont {Phark}}]{Wang_2023}%
      \BibitemOpen
      \bibfield  {author} {\bibinfo {author} {\bibfnamefont {Y.}~\bibnamefont
      {Wang}}, \bibinfo {author} {\bibfnamefont {M.}~\bibnamefont {Haze}}, \bibinfo
      {author} {\bibfnamefont {H.}~\bibnamefont {Bui}}, \bibinfo {author}
      {\bibfnamefont {W.-h.}\ \bibnamefont {Soe}}, \bibinfo {author} {\bibfnamefont
      {H.}~\bibnamefont {Aubin}}, \bibinfo {author} {\bibfnamefont
      {A.}~\bibnamefont {Ardavan}}, \bibinfo {author} {\bibfnamefont
      {A.}~\bibnamefont {Heinrich}},\ and\ \bibinfo {author} {\bibfnamefont
      {S.}~\bibnamefont {Phark}},\ }\bibfield  {title} {\bibinfo {title} {Universal
      quantum control of an atomic spin qubit on a surface},\ }\href
      {https://doi.org/10.1038/s41534-023-00716-6} {\bibfield  {journal} {\bibinfo
      {journal} {npj Quantum Inf.}\ }\textbf {\bibinfo {volume} {9}},\ \bibinfo
      {pages} {48} (\bibinfo {year} {2023})}\BibitemShut {NoStop}%
    \bibitem [{\citenamefont {Bedow}\ \emph {et~al.}(2022)\citenamefont {Bedow},
      \citenamefont {Mascot},\ and\ \citenamefont {Morr}}]{Bedow_2022}%
      \BibitemOpen
      \bibfield  {author} {\bibinfo {author} {\bibfnamefont {J.}~\bibnamefont
      {Bedow}}, \bibinfo {author} {\bibfnamefont {E.}~\bibnamefont {Mascot}},\ and\
      \bibinfo {author} {\bibfnamefont {D.~K.}\ \bibnamefont {Morr}},\ }\bibfield
      {title} {\bibinfo {title} {Emergence and manipulation of non-equilibrium
      {{Yu-Shiba-Rusinov}} states},\ }\href
      {https://doi.org/10.1038/s42005-022-01050-7} {\bibfield  {journal} {\bibinfo
      {journal} {Commun. Phys.}\ }\textbf {\bibinfo {volume} {5}},\ \bibinfo
      {pages} {281} (\bibinfo {year} {2022})}\BibitemShut {NoStop}%
    \bibitem [{\citenamefont {Balatsky}\ \emph {et~al.}(2006)\citenamefont
      {Balatsky}, \citenamefont {Vekhter},\ and\ \citenamefont
      {Zhu}}]{Balatsky_2006}%
      \BibitemOpen
      \bibfield  {author} {\bibinfo {author} {\bibfnamefont {A.~V.}\ \bibnamefont
      {Balatsky}}, \bibinfo {author} {\bibfnamefont {I.}~\bibnamefont {Vekhter}},\
      and\ \bibinfo {author} {\bibfnamefont {J.-X.}\ \bibnamefont {Zhu}},\
      }\bibfield  {title} {\bibinfo {title} {Impurity-induced states in
      conventional and unconventional superconductors},\ }\href
      {https://doi.org/10.1103/RevModPhys.78.373} {\bibfield  {journal} {\bibinfo
      {journal} {Rev. Mod. Phys.}\ }\textbf {\bibinfo {volume} {78}},\ \bibinfo
      {pages} {373} (\bibinfo {year} {2006})}\BibitemShut {NoStop}%
    \bibitem [{\citenamefont {Heinrich}\ \emph {et~al.}(2018)\citenamefont
      {Heinrich}, \citenamefont {Pascual},\ and\ \citenamefont
      {Franke}}]{Heinrich_2018}%
      \BibitemOpen
      \bibfield  {author} {\bibinfo {author} {\bibfnamefont {B.~W.}\ \bibnamefont
      {Heinrich}}, \bibinfo {author} {\bibfnamefont {J.~I.}\ \bibnamefont
      {Pascual}},\ and\ \bibinfo {author} {\bibfnamefont {K.~J.}\ \bibnamefont
      {Franke}},\ }\bibfield  {title} {\bibinfo {title} {Single magnetic adsorbates
      on s -{{Wave}} superconductors},\ }\href
      {https://doi.org/10.1016/j.progsurf.2018.01.001} {\bibfield  {journal}
      {\bibinfo  {journal} {Prog. Surf. Sci.}\ }\textbf {\bibinfo {volume} {93}},\
      \bibinfo {pages} {1} (\bibinfo {year} {2018})}\BibitemShut {NoStop}%
    \bibitem [{\citenamefont {Sakurai}(1970)}]{Sakurai_1970}%
      \BibitemOpen
      \bibfield  {author} {\bibinfo {author} {\bibfnamefont {A.}~\bibnamefont
      {Sakurai}},\ }\bibfield  {title} {\bibinfo {title} {Comments on
      {{Superconductors}} with {{Magnetic Impurities}}},\ }\href
      {https://doi.org/10.1143/PTP.44.1472} {\bibfield  {journal} {\bibinfo
      {journal} {Prog. Theor. Phys.}\ }\textbf {\bibinfo {volume} {44}},\ \bibinfo
      {pages} {1472} (\bibinfo {year} {1970})}\BibitemShut {NoStop}%
    \bibitem [{\citenamefont {Salkola}\ \emph {et~al.}(1997)\citenamefont
      {Salkola}, \citenamefont {Balatsky},\ and\ \citenamefont
      {Schrieffer}}]{Salkola_1997}%
      \BibitemOpen
      \bibfield  {author} {\bibinfo {author} {\bibfnamefont {M.~I.}\ \bibnamefont
      {Salkola}}, \bibinfo {author} {\bibfnamefont {A.~V.}\ \bibnamefont
      {Balatsky}},\ and\ \bibinfo {author} {\bibfnamefont {J.~R.}\ \bibnamefont
      {Schrieffer}},\ }\bibfield  {title} {\bibinfo {title} {Spectral properties of
      quasiparticle excitations induced by magnetic moments in superconductors},\
      }\href {https://doi.org/10.1103/PhysRevB.55.12648} {\bibfield  {journal}
      {\bibinfo  {journal} {Phys. Rev. B}\ }\textbf {\bibinfo {volume} {55}},\
      \bibinfo {pages} {12648} (\bibinfo {year} {1997})}\BibitemShut {NoStop}%
    \bibitem [{\citenamefont {Bazaliy}\ and\ \citenamefont
      {Jones}(2000)}]{Bazaliy_2000}%
      \BibitemOpen
      \bibfield  {author} {\bibinfo {author} {\bibfnamefont {Y.~B.}\ \bibnamefont
      {Bazaliy}}\ and\ \bibinfo {author} {\bibfnamefont {B.~A.}\ \bibnamefont
      {Jones}},\ }\bibfield  {title} {\bibinfo {title} {Magnetic impurity in a
      superconductor: Local phase transitions and finite size effects},\ }\href
      {https://doi.org/10.1063/1.373404} {\bibfield  {journal} {\bibinfo  {journal}
      {J. Appl. Phys.}\ }\textbf {\bibinfo {volume} {87}},\ \bibinfo {pages} {5561}
      (\bibinfo {year} {2000})}\BibitemShut {NoStop}%
    \bibitem [{\citenamefont {Morr}\ and\ \citenamefont
      {Stavropoulos}(2003)}]{Morr_2003a}%
      \BibitemOpen
      \bibfield  {author} {\bibinfo {author} {\bibfnamefont {D.~K.}\ \bibnamefont
      {Morr}}\ and\ \bibinfo {author} {\bibfnamefont {N.~A.}\ \bibnamefont
      {Stavropoulos}},\ }\bibfield  {title} {\bibinfo {title} {Quantum interference
      between impurities: {{Creating}} novel many-body states in {\emph{s}} -wave
      superconductors},\ }\href {https://doi.org/10.1103/PhysRevB.67.020502}
      {\bibfield  {journal} {\bibinfo  {journal} {Phys. Rev. B}\ }\textbf {\bibinfo
      {volume} {67}},\ \bibinfo {pages} {020502} (\bibinfo {year}
      {2003})}\BibitemShut {NoStop}%
    \bibitem [{\citenamefont {Morr}\ and\ \citenamefont {Yoon}(2006)}]{Morr_2006a}%
      \BibitemOpen
      \bibfield  {author} {\bibinfo {author} {\bibfnamefont {D.~K.}\ \bibnamefont
      {Morr}}\ and\ \bibinfo {author} {\bibfnamefont {J.}~\bibnamefont {Yoon}},\
      }\bibfield  {title} {\bibinfo {title} {Impurities, quantum interference, and
      quantum phase transitions in s -wave superconductors},\ }\href
      {https://doi.org/10.1103/PhysRevB.73.224511} {\bibfield  {journal} {\bibinfo
      {journal} {Phys. Rev. B}\ }\textbf {\bibinfo {volume} {73}},\ \bibinfo
      {pages} {224511} (\bibinfo {year} {2006})}\BibitemShut {NoStop}%
    \bibitem [{\citenamefont {Raussendorf}\ and\ \citenamefont
      {Harrington}(2007)}]{Raussendorf_2007}%
      \BibitemOpen
      \bibfield  {author} {\bibinfo {author} {\bibfnamefont {R.}~\bibnamefont
      {Raussendorf}}\ and\ \bibinfo {author} {\bibfnamefont {J.}~\bibnamefont
      {Harrington}},\ }\bibfield  {title} {\bibinfo {title} {Fault-tolerant quantum
      computation with high threshold in two dimensions},\ }\href
      {https://doi.org/10.1103/PhysRevLett.98.190504} {\bibfield  {journal}
      {\bibinfo  {journal} {Phys. Rev. Lett.}\ }\textbf {\bibinfo {volume} {98}},\
      \bibinfo {pages} {190504} (\bibinfo {year} {2007})}\BibitemShut {NoStop}%
    \bibitem [{\citenamefont {Stace}\ \emph {et~al.}(2009)\citenamefont {Stace},
      \citenamefont {Barrett},\ and\ \citenamefont {Doherty}}]{Stace_2009}%
      \BibitemOpen
      \bibfield  {author} {\bibinfo {author} {\bibfnamefont {T.~M.}\ \bibnamefont
      {Stace}}, \bibinfo {author} {\bibfnamefont {S.~D.}\ \bibnamefont {Barrett}},\
      and\ \bibinfo {author} {\bibfnamefont {A.~C.}\ \bibnamefont {Doherty}},\
      }\bibfield  {title} {\bibinfo {title} {Thresholds for topological codes in
      the presence of loss},\ }\href
      {https://doi.org/10.1103/PhysRevLett.102.200501} {\bibfield  {journal}
      {\bibinfo  {journal} {Phys. Rev. Lett.}\ }\textbf {\bibinfo {volume} {102}},\
      \bibinfo {pages} {200501} (\bibinfo {year} {2009})}\BibitemShut {NoStop}%
    \bibitem [{\citenamefont {Ring}\ and\ \citenamefont
      {Schuck}(1980)}]{Ring_1980}%
      \BibitemOpen
      \bibfield  {author} {\bibinfo {author} {\bibfnamefont {P.}~\bibnamefont
      {Ring}}\ and\ \bibinfo {author} {\bibfnamefont {P.}~\bibnamefont {Schuck}},\
      }\href@noop {} {\emph {\bibinfo {title} {The {{Nuclear Many-Body
      Problem}}}}},\ \bibinfo {edition} {1st}\ ed.,\ \bibinfo {number} {1864-5879}\
      (\bibinfo  {publisher} {{Springer Berlin, Heidelberg}},\ \bibinfo {year}
      {1980})\BibitemShut {NoStop}%
    \bibitem [{\citenamefont {Onishi}\ and\ \citenamefont
      {Yoshida}(1966)}]{Onishi_1966}%
      \BibitemOpen
      \bibfield  {author} {\bibinfo {author} {\bibfnamefont {N.}~\bibnamefont
      {Onishi}}\ and\ \bibinfo {author} {\bibfnamefont {S.}~\bibnamefont
      {Yoshida}},\ }\bibfield  {title} {\bibinfo {title} {Generator coordinate
      method applied to nuclei in the transition region},\ }\href
      {https://doi.org/https://doi.org/10.1016/0029-5582(66)90096-4} {\bibfield
      {journal} {\bibinfo  {journal} {Nucl. Phys.}\ }\textbf {\bibinfo {volume}
      {80}},\ \bibinfo {pages} {367} (\bibinfo {year} {1966})}\BibitemShut
      {NoStop}%
\end{thebibliography}

%

\end{document}